\documentclass[]{article}
\usepackage{emulateapj}
\usepackage{psfig}
\usepackage{graphicx}

\advance \voffset by 1.00cm\relax
\def\msun{{\,M_{\odot}}}

\def\kms{{\,{\rm km}\,{\rm s}^{-1}}}
\begin{document}

\title{Initial Populations of Black Holes in Star Clusters}

\author{Krzysztof Belczynski\altaffilmark{1,2} Aleksander Sadowski\altaffilmark{3}, 
        Frederic A.\ Rasio\altaffilmark{4}, Tomasz Bulik \altaffilmark{5}}

\affil{
     $^{1}$ New Mexico State University, Dept of Astronomy,
       1320 Fregner Mall, Las Cruces, NM 88003\\
     $^{2}$ Tombaugh Fellow\\
     $^{3}$ Astronomical Observatory, Warsaw University,
            Al.\ Ujazdowskie 4, 00-478, Warsaw, Poland \\
     $^{4}$ Northwestern University, Dept of Physics and Astronomy, 
       2145 Sheridan Rd, Evanston, IL 60208\\
     $^{5}$ Nicolaus Copernicus Astronomical Center,
            Bartycka 18, 00-716 Warszawa, Poland\\
     kbelczyn@nmsu.edu, oleks@camk.edu.pl, rasio@northwestern.edu, 
     bulik@camk.edu.pl }

\begin{abstract} 

Using an updated population synthesis code we study the formation and
evolution of black holes (BHs) in young star clusters following a massive starburst.
This study continues and improves on the initial work described by Belczynski, 
Sadowski \& Rasio (2004). In our new calculations we account for the possible ejections 
of BHs and their progenitors from clusters because of natal kicks imparted by 
supernovae and recoil following binary disruptions. The results indicate that 
the properties of both retained BHs in clusters and ejected BHs (forming a
field population) depend sensitively on the depth of the cluster potential. 
In particular, most BHs ejected from binaries are also ejected from clusters
with central escape speeds $V_{\rm esc} \la 100\,{\rm km}\,{\rm s}^{-1}$. 
Conversely, most BHs remaining in binaries are retained by clusters with 
$V_{\rm esc} \ga 50\,{\rm km}\,{\rm s}^{-1}$. 
BHs from single star evolution are also affected significantly: about half of 
the BHs originating from primordial single stars are ejected from clusters with 
$V_{\rm esc}\la 50\,{\rm km}\,{\rm s}^{-1}$.
Our results lay a foundation for theoretical studies of the formation of BH X-ray 
binaries in and around star clusters, including possible ``ultra-luminous'' sources, 
as well as merging BH--BH 
binaries detectable with future gravitational-wave observatories.
\end{abstract}


\section{INTRODUCTION}

\subsection{Black Holes in Star Clusters}

Theoretical arguments and many observations suggest
that BHs should form in significant numbers in star clusters.
Simple assumptions about the stellar initial mass function
(IMF) and stellar evolution indicate that out of $N$ stars formed
initially, $\sim 10^{-4}-10^{-3}\,N$ should produce BHs as remnants
after $\sim 20\,$Myr. Thus any star cluster containing
initially more than $\sim 10^4$ stars should contain at least some BHs;
large super star clusters and globular clusters should have formed
many hundreds of BHs initially, and even larger systems such as
galactic nuclei may contain many thousands to tens of thousands.
 
Not surprisingly, observations are most sensitive to (and have
provided constraints mainly on) the most massive BHs that may be
present in the cores of very dense clusters (van der Marel 2004). 
For example, recent observations and dynamical modeling of the globular clusters
M15 and G1 indicate the presence of a central BH with a 
mass $\sim 10^3-10^4\,M_\odot$ (Gerssen et al.\ 2002, 2003;
Gebhardt et al.\ 2002, 2005). However, direct $N$-body simulations by
Baumgardt et al.\ (2003a,b) suggest that the observations of
M15 and G1, and, in general, the properties of all {\em core-collapsed\/} clusters, 
could be explained equally well by the presence of many compact
remnants (heavy white dwarfs, neutron stars, or $\sim 3-15\msun$ BHs)
near the center without a massive BH (cf.\ van der Marel 2004; Gebhardt et al.\ 2005). 
On the other hand,
$N$-body simulations also suggest that many {\em non-core-collapsed\/} clusters (representing about
80\% of globular clusters in the Milky Way) could contain central 
massive BH (Baumgardt et al.\ 2004, 2005).

In any case, when the correlation between central BH mass and bulge mass
in galaxies (e.g., H\"aring \& Rix 2004) is extrapolated to smaller stellar 
systems like globular 
clusters, the inferred BH masses are indeed $\sim 10^3-10^4\,M_\odot$. 
These are much larger than a canonical $\sim10\,M_\odot$ stellar-mass 
BH (see, however, \S 3.1.6), but much smaller than the $\sim10^6-10^9\,M_\odot$ 
of supermassive BHs. Hence, these objects are
often called {\em intermediate-mass black holes\/} (IMBHs; see, e.g., Miller \& 
Colbert 2004). 

Further observational evidence for IMBHs in dense star clusters comes from many recent 
{\em Chandra\/} and XMM-{\em Newton\/} observations of ``ultra-luminous'' X-ray sources 
(ULXs), which are often (although not always) clearly associated with young star clusters 
and whose high X-ray luminosities in many cases suggest a compact object
mass of at least $\sim 10^2\,M_\odot$ (Cropper et al.\ 2004; Ebisuzaki et al.\ 2001;
Kaaret et al.\ 2001; Miller et al.\ 2003). In many cases, however,
beamed emission by an accreting stellar-mass BH may provide an alternative explanation 
(King et al.\ 2001; King 2004; Zezas \& Fabbiano 2002).

One natural path to the formation of a massive object at the center of 
any young stellar system with
a high enough density is through runaway collisions and mergers of
massive stars following gravothermal contraction and core collapse 
(Ebisuzaki et al.\ 2001;
Portegies Zwart \& McMillan 2002; G\"urkan, Freitag, \& Rasio 2004). 
These runaways occur when massive stars can drive core collapse {\em before 
they evolve\/}. Alternatively, if the most massive stars in the cluster are 
allowed to evolve and produce supernovae, the gravothermal contraction of 
the cluster will be reversed by the sudden mass loss, and 
many stellar-mass BHs will be formed. 

The final fate of a cluster with a significant component of
stellar-mass BHs remains highly uncertain. This is because realistic
dynamical simulations for such clusters (containing a large number of
BHs {\em and\/} ordinary stars with a realistic mass spectrum) 
have yet to be performed. For old and relatively small systems (such as small
globular clusters), complete evaporation is likely (with essentially all the
stellar-mass BHs ejected from the cluster through three-body and four-body
interactions in the dense core). This is expected 
theoretically on the basis of simple qualitative arguments based
on Spitzer's ``mass-segregation instability'' applied to BHs (Kulkarni et al.\ 1993; 
Sigurdsson \& Hernquist 1993; Watters et al.\ 2000)
and has been demonstrated by dynamical simulations (Portegies Zwart \& 
McMillan 2000; O'Leary et al.\ 2005). However, it has been
suggested that, if stellar-mass BHs are formed with a relatively broad mass spectrum (a
likely outcome for stars of very low metallicity; see Heger et al.\ 2003), the
most massive BH could resist ejection, even from a cluster with low escape 
velocity. These more massive BHs could then grow
by repeatedly forming binaries (through exchange interactions)
with other BHs and merging with their companions (Miller \& Hamilton 2002;
G\"ultekin, Miller, \& Hamilton 2004). However, as 
most interactions will probably result in the ejection 
of one of the lighter BHs, it is unclear whether any object could grow 
substantially through this mechanism before running out of companions to merge
with. A single stellar-mass BH remaining at the center of a globular cluster
is very unlikely to become detectable as an X-ray binary (Kalogera, King, \& Rasio 2004).

In addition to its obvious relevance to X-ray astronomy, the dynamics of
BHs in clusters also plays an important role in the 
theoretical modeling of gravitational-wave (GW) sources and the development of 
data analysis and detection strategies for these sources. In particular, 
the growth of a massive BH by repeated mergers of stellar-mass BHs spiraling into an
IMBH at the center of a dense star cluster may provide an important
source of low-frequency GWs for LISA, the Laser-Interferometer Space Antenna
(Miller 2002; Will 2004). Similarly, dynamical hardening and ejections of binaries 
from dense clusters of stellar-mass BHs could lead to greatly enhanced
rates of BH--BH mergers detectable by LIGO and other ground-based
interferometers (Portegies Zwart \& McMillan 2000; O'Leary et al.\ 2005).

A crucial starting point for any detailed study of BHs in clusters is an
accurate description of the initial BH population. Here,
``initial'' means on a timescale short compared to the later dynamical
evolution timescale. Indeed most $N$-body simulations of star cluster dynamics
never attempt to model the brief, initial phase of rapid massive star evolution
leading to BH formation. The goal of our work here is to provide the most
up-to-date and detailed description of these initial BH populations. This means
that we must compute the evolution of a large number of massive stars, including
a large fraction of binaries, all the way to BH formation, i.e., on a timescale
$\sim 10-100\,$Myr, taking into account a variety of possible cluster environments.

\subsection{Previous Work}

In a previous study (Belczynski et al. 2004; hereafter Paper~I)
we studied young populations of BHs formed in a massive
starburst, without explicitly taking into account that most stars are formed
in clusters. For many representative models we computed the numbers
of BHs, both single and in various types of binaries, at various ages,
as well as the physical properties of different systems (e.g., binary period 
and BH mass distributions). We also discussed in detail the evolutionary
channels responsible for these properties. 

In this follow-up study, we consider the possible ejection of these BHs from
star clusters with different escape speeds, taking into account the recoil
imparted by supernovae
(SNe) and binary disruptions. During SNe, mass loss and any asymmetry in the
explosion (e.g., in neutrino emission) can impart large extra speeds to newly 
formed compact objects. If a compact object is formed in a binary system, 
the binary may either {\em (i)} survive the explosion, but its orbital
parameters are changed and the system (center-of-mass) speed changes,
 or {\em (ii)} the
binary is disrupted and the newly formed compact object and its companion 
fly apart on separate trajectories. The secondary star in a binary may later undergo a
SN explosion as well, provided that it is massive enough. The effects of this 
second explosion are equally important in determining the final characteristics
of compact objects.

In Paper~I we included the effects of SNe, both natal kicks 
and mass loss, on the formation and evolution of BHs (single and in binaries), 
but we did not keep track of which BHs and binaries would be retained in
their parent cluster. 
Starbursts form most of their stars in dense clusters with a broad range of masses
and central potentials (and hence escape speeds; see, e.g., Elmegreen et al.\
2002; McCrady et al.\ 2003; Melo et al.\ 2005). Smaller clusters of 
$\sim 10^4\,M_\odot$ (open clusters or ``young populous clusters,'' such as the 
Arches and Quintuplet clusters in our Galactic center) could have escape speeds 
as low as $V_{\rm esc} \la 10\,{\rm km}\,{\rm s}^{-1}$ while the 
largest ``super star clusters'' with much deeper potential wells could 
have $V_{\rm esc} \ga 100\,{\rm km}\,{\rm s}^{-1}$.
On the other hand the natal kick velocities could be relatively high,
$\sim 100-500\,{\rm km}\,{\rm s}^{-1}$ for low-mass BHs,
so that a large fraction of BHs might leave the
cluster early in the evolution. 

Here we repeat our study of young BH populations taking into account
ejections from star clusters.
We perform our calculations with a slightly updated version of our
population synthesis code {\tt StarTrack} (\S2) and we
present results for both the retained cluster BH populations and the 
ejected BHs, which will eventually become part of the field BH population
surrounding the surviving clusters.   
Our models and assumptions are discussed in \S2, with particular emphasis on
the updates since Paper~I. In \S3 we present our new results and in
\S4 we provide a summary and discussion.

\section{MODEL DESCRIPTION AND ASSUMPTIONS}

\subsection {Population Synthesis Code}

Our investigation is based on a standard population synthesis method.
We use the {\tt StarTrack} code (Belczynski, Kalogera \& Bulik 2002, 
hereafter BKB02), which has been revised and improved significantly 
over the past few years (Belczynski et al.\ 2006). Our calculations do 
not include any treatment of dynamical interactions (collisions) between 
binaries and single stars or other binaries\footnote{The only exception 
is that we take into account implicitly the likely mass segregation of 
massive stars into the cluster core. See \S2.3.}. In particular, the 
star clusters we consider are assumed to have avoided the `runaway
collision instability'' that can drive rapid collisions and mergers of
massive main-sequence stars during an early episode of cluster core
collapse (Freitag et al.\ 2006a,b). Instead, our results can provide 
highly realistic initial conditions for dynamical simulations of dense 
star clusters in which the early phase of massive star evolution proceeded 
`normally,'' without significant influence from cluster dynamics.

All stars are evolved based on the metallicity- and wind-mass-loss-dependent 
models of Hurley, Pols \& Tout (2002), with a few improvements described in 
BKB02. The main code parameters we use correspond to the standard model presented 
in \S2 of BKB02 and are also described in Paper~I. Each star, 
either single or a binary component, is placed initially on the
zero-age main sequence (ZAMS) and then evolved 
through a sequence of distinct phases: main sequence (MS), Hertzsprung Gap 
(HG), red giant branch (RG), core He burning (CHeB), asymptotic giant branch 
(AGB); if a star gets stripped of its H-rich envelope, either through wind 
mass loss or Roche lobe overflow (RLOF) it becomes a naked helium star (He). 
The nuclear evolution 
leads ultimately to the formation of a compact object. Depending on the pre-collapse 
mass and initial composition this may be a white dwarf (WD), a neutron star 
(NS) or a BH. 

The population synthesis code allows us to study the evolution of both
single and binary stars. Binary star components are evolved as single stars 
while no interactions are taking place. We model the following processes, which can 
alter the binary
orbit and subsequent evolution of the components: tidal
interactions, magnetic braking, gravitational radiation, and angular 
momentum changes due to mass loss. Binary components may interact through
mass transfer and accretion phases. We take into account various modes of mass
transfer: wind accretion and RLOF; conservative and non-conservative; 
stable or dynamically unstable (leading to common-envelope evolution). 
The mass transfer rates are calculated from the specific binary
configurations and physical properties (masses, evolutionary stages, etc.) of 
the stars involved. Binary components may loose or gain
mass, while the binary orbit may either expand or shrink in response. Moreover,
we allow for binary mergers driven by orbital decay. In this study, 
we evolve binary merger products assuming that they restart on the ZAMS.
An exception is made when a BH takes part in the merger, in which case 
we assume the remnant object to be a BH again.
The mass of the merger product is assumed equal to the total parent binary mass for 
unevolved and compact remnant components; however we assume complete envelope mass 
loss from any evolved star (HG, RG, CHeB, or evolved He star) involved in a merger. 

A few additions and updates to {\tt StarTrack} since Paper~I are worth mentioning
here (see Belczynski et al.\ 2006 for more details). 
System velocities are now tracked for all stars (single and binaries) after SNe (see 
\S 2.3). The new magnetic braking law of Ivanova \& Taam (2003) has been adopted, 
although this has minimal impact on our results for BHs.  Two new types of WDs have been
introduced: hydrogen and ``hybrid'' (these are possible BH donors in binaries).
An improved criterion is adopted for CHeB stars to discriminate between those with 
convective ($M < 7\msun$) and radiative ($M \geq 7\msun$) envelopes; this affects the 
stellar response to mass loss. We have also added a new tidal term for RLOF rate 
calculations. 
 
Some minor problems in the calculations of Paper~I were also identified and
are corrected in this study. The evolution of a small fraction of BH RLOF systems 
with donors at the end of the RG
stage was terminated when the donor contracted and detached after 
entering a CHeB phase. However, the donor may restart RLOF during 
expansion on the AGB, which is now properly accounted for.   
Another small fraction of systems, evolving through the rapid RLOF phase
with HG donors, were previously classified as mergers and subsequently
evolved as single
stars (merger products). However, the RLOF at that stage may be dynamically
stable and in some cases a binary system may survive and continue its
evolution, which is now also properly taken into account. None of these corrections
affect the results of Paper~I significantly.

\subsection {Black Hole Formation}

Black holes originate from the most massive stars. The formation time is
calculated for each star using the stellar models of Hurley et al.\ (2000) and
Woosley (1986). For intermediate-mass stars the FeNi core collapses and forms
a hot proto-NS or a low-mass BH. Part of the envelope falls back onto the central
object while the rest is assumed to be ejected in a SN explosion. We
use the results of Fryer (1999) and Fryer \& Kalogera (2001) to determine how
much matter is ejected. In general, for the highest masses ($>30\,M_{\odot}$ 
for low-metallicity models) total fall-back is expected, with no accompanying SN
explosion. 

Motivated by the large observed velocities of radio pulsars we assume significant
asymmetries in SN explosions. Here we adopt the 
kick velocity distribution of Arzoumanian, Cordes \& Chernoff (2002), taking into 
consideration more recent observations (e.g., White \& Van Paradijs 1996; Mirabel 
\& Rodrigues 2003). NSs receive full kicks drawn
from the bimodal distribution of Arzoumanian et al.\ (2002). Many BHs form
through partial fall back of material initially expelled in a SN explosion, but 
then accreted back onto the central BH. For these the kick velocity is lowered
proportionally to the mass of accreted material (for details see BKB02). 
For the most massive stars, the BH forms silently through a direct collapse without 
accompanying SN explosion, and in this case we assume {\em no\/} BH natal 
kick. The mass loss and kick velocity together determine whether a binary hosting 
the BH progenitor is disrupted by the SN explosion. 

Our calculated initial-to-final mass relation for various metallicities is
discussed in detail in Paper~I, where it is also demonstrated that (within our BH kick model)
for solar metallicity many BHs are formed with lowered kicks through fall back. This 
occurs for single stars with initial masses in the range $20 - 42 M_{\odot}$ and 
$50 - 70 M_{\odot}$. For metallicity $Z = 0.001$, BHs receive a kick in the narrower 
ranges $18 - 25 M_{\odot}$ and $39 - 54 M_{\odot}$, while for $Z = 0.0001$ only BHs 
formed from stars of $18 - 24 M_{\odot}$ receive kicks, with others forming silently.

\subsection{Spatial Velocities} 

All stars, single and binaries, are assumed to have zero initial
velocities. This means we are neglecting their orbital speeds
within the cluster. Indeed, for a variety of reasons (e.g., relaxation
toward energy equipartition, formation near the cluster center),
massive stars (BH progenitors) are expected to have
lower velocity dispersions than the cluster average, itself
much lower than the escape speed from the cluster center. We now discuss
how to compute the velocities acquired by single stars, binaries, or their disrupted
components, following a SN explosion.
For disrupted binary components this is, to our knowledge, the first derivation of 
such results allowing for initially {\em eccentric\/} orbits.

{\em Surviving Binaries.\/} For each massive star, the time of the SN event 
is set by the single star models (taking into account mass variations 
due to winds and binary interactions).  When either component of a binary
reaches this stage, we generate a random location in the orbit for the
event to take place (note that for eccentric binaries this choice will
affect the outcome, since the instantaneous separation and relative
velocities are different at different locations along the orbit).  The core
collapse event is assumed to be instantaneous and the mass of the
remnant is calculated as in Paper~I. Note that if the remnant is
formed through complete fall-back (leading always to direct BH
formation), we do not expect a SN explosion (hence no kick and no mass 
loss) and the orbit remains unchanged (Fryer 1999). When a BH is formed 
through partial fall-back we treat the event as a SN explosion (see 
Podsiadlowski et al.\ 2002). 

We calculate the effect of a SN event on binaries in three steps. First,
we estimate the mass of the remnant. The rest of the exploding star is
immediately lost from the binary (with the specific angular momentum of the
exploding component). We assume that the ejecta do not have
any effect on the companion. Second, we calculate the
compact object velocity, which is the vector sum of the orbital velocity
of the pre-collapse star at the random orbital position and the kick velocity.
The kick velocity is assumed to be randomly oriented and its magnitude
is drawn from our assumed distribution. The kick magnitude is
also scaled by the amount of material ejected in the SN explosion,
 \begin{equation} 
 w=(1-f_{\rm fb})*V, 
 \end{equation} 
 where $V$ is the kick magnitude drawn from the assumed
distribution, $f_{\rm fb}$ is a fall back parameter (for details 
see Paper~I), and $w$ is the kick magnitude we use in our
calculations. For NS remnants with no fall-back ($f_{\rm fb}=0$),
$w=V$. In our standard model we use a kick magnitude distribution derived 
by Arzoumanian et al. (2002): a weighted sum 
of two Gaussians, one with $\sigma=90$\ km~s$^{-1}$ (40\%) and the second 
with $\sigma=500$~km~s$^{-1}$ (60\%). In the third step, we calculate the 
total energy (potential and kinetic) of the new
orbit for the remnant (new velocity and mass, same relative position)
and its companion. If the total energy is positive, then the system is
disrupted, and its components will evolve separately. We calculate their
subsequent evolution as single stars, and their trajectories are followed 
(see below). If the total binary energy is
negative, the system remains bound after the SN explosion, and we calculate its
new parameters ($e$ and $A$). We also check whether the two components have
merged following the SN mass loss and kick, in which case we the evolution
of a merger is followed (see \S\,2.1). These type of mergers are extremely
rare and they do not affect our results. Finally, we calculate the post-SN 
center of mass velocity of the binary.

{\em Disrupted Binaries.\/} The velocities of the stellar components after a 
system has been disrupted have been calculated by Tauris \& Takens (1998) for 
the case of a circular pre-SN orbit. In this paper we take a more general 
approach where the pre-SN orbit can have an arbitrary eccentricity. 
We begin with the coordinate system (I): the center of mass (CM) coordinate system 
before the SN explosion. At the time of the SN explosion the velocities of the 
two stars are
\begin{equation}
{\vec v}^I_1={-M_2 \vec v \over M_1+M_2} 
\end{equation}
\begin{equation}
{\vec v}^I_2={M_1 \vec v  \over M_1+M_2} 
\end{equation}
where $\vec v$ is the relative velocity 
and a superscript $I$ indicates the coordinate system we use.
The separation vector between the stars on the orbit 
at the moment of the SN explosion is $r_0 \vec n$.
The SN explosion introduces a kick $\vec w$ 
on the newly formed compact star and leads
to the ejection of a shell. Thus after the SN
explosion star~1 has mass $M_{1f}$, and its velocity
in the pre-SN CM coordinate system (I) is
\begin{equation}
 {\vec v}^I_{1i}={\vec v}^I_1+\vec w .
 \end{equation}
The expanding shell (with velocity ${\vec v}_{im}$) hits the companion. 
The effects of the impact have been estimated by Wheeler, Lecar and McKee 
(1975), but it was shown (Kalogera 1996) that they are not large. 
The velocity of the companion after the expanding shell
decouples from the system is 
\begin{equation}
 {\vec v}^I_{2i}={\vec v}^I_2+{\vec v}_{im}.
 \end{equation}
In most cases $v_{im}$ is small, and we neglect it
in our treatment of the orbits. 
The direction between the stars remains the same, ${\vec n}^{II}={\vec n}^{I}$.

Here we also assume that the shell velocity  satisfies
$ v_{im} \gg r_0/P$
where $P$ is the orbital period of the pre-SN system, so 
that the shell decouples from the binary instantaneously.

We now introduce a second coordinate system (II): the CM
system of the two remaining stars after the explosion. 
This system moves with the velocity 
\begin{equation}
 {\vec v}^{II}_{CM} = { M_{1f}{\vec v}^I_{1i}+ M_2 {\vec v}^I_{2i} \over M_{1f}+ M_2} 
 \end{equation}
with respect to system (I). 
The relative velocity 
 of the two stars (the newly formed
compact object and the companion) in this system is
\begin{equation}
{\vec v}^{II} = {\vec v}+{\vec w}-{\vec v}_{im}
\end{equation}
after shell decoupling.
The angular momentum of  the two stars   is 
\begin{equation}
 J = \mu r_0 {\vec n}^{II} \times {\vec v}^{II},
 \end{equation}
where ${\vec n}^{II}={\vec n}^{I}$
It is convenient now to introduce the coordinate system  (III) in which 
the angular momentum is aligned with the $z$-axis. 
The transformation between (II) and (III) is a simple rotation,
which we denote as $\cal R$: $v^{III}={\cal R}v^{II}$,$n^{III}={\cal R}n^{II}$ .
In the  coordinate system (III) the 
two stars move on a hyperbolic orbit described by 
\begin{equation}
 r = {p \over 1 + \epsilon \cos\phi} \label{traj}
\end{equation}
 where 
\begin{eqnarray}
p={J^2\over \alpha\mu} \ \ {\rm and} \ \
\epsilon=\sqrt{1+{2EJ^2\over\alpha^2\mu}},
\end{eqnarray}
$E=\mu (v^{II})^2/2-\alpha/|r_0|$ is the (positive) 
energy of the system,  $\alpha=G M_{1f}M_{2}$, and 
$\mu = M_{1f}M_2/(M_{1f}+M_2)$ is the reduced mass.
Using the conservation of energy we find the value 
of the final relative velocity in (III):
\begin{equation} |v^{III}_f| = \sqrt{2E\over \mu}.
\end{equation}
It follows from the conservation of angular momentum 
 that  the relative velocity ${\vec v}^{III}_f$ at infinity 
is parallel to the separation vector 
 ${\vec n}^{III}_f$. The motion  is confined to the $x-y$ plane in (III), 
 so what remains now is to find the angle between
${\vec n}^{III}$ and ${\vec n}^{III}_f$.
This can easily be found using equation \ref{traj}. We first find the
initial position of the star on the trajectory $\varphi_i$ (see Fig.~1),
\begin{equation}
 \cos\varphi_i= {1\over\epsilon}\left({p\over r_0 - 1 }\right) .
 \end{equation}
The sign of the sine of $\varphi_i$ depends on whether the stars are initially
on the ascending or descending branch of the hyperbola. In the 
ascending branch the scalar product  $\vec v^{III}_i \vec r^{III}_0<0$, thus 
\begin{equation}
 \sin\varphi_i = {\rm sgn} (\vec v^{III}_i 
 \vec r^{III}_0)\sqrt{1-\cos^2\phi_f}
 \end{equation}
The final  position on the orbit is given by
\begin{equation}
 \cos\varphi_f=-{1\over \epsilon}.
 \end{equation}
Finally we see from Fig.~1 that   $\sin\varphi_f>0$.

With these results we can calculate the direction between the stars at 
$r=\infty$: $n^{III}_f=T(\varphi_f-\varphi_i)n^{III}$, where
$T(\phi)$ is the matrix of rotation around the z axis.
We now have the result that the relative final velocity 
in the coordinate system (III) is 
\begin{equation}
v^{III}_f=\sqrt{2E\over \mu} n^{III}_f.
\end{equation}
We find the velocities of the individual stars in (I) by reversing the path 
of transformations we followed above,
\begin{eqnarray}
v^{I}_{1f}={\cal R}^{-1}\left({-M_2 v^{III}_f\over M_{1f}+M2}\right) + v^{II}_{CM}\\
v^{I}_{2f}={\cal R}^{-1}\left({M_{1f} v^{III}_f\over M_{1f}+M2}\right) + v^{II}_{CM}
\end{eqnarray}
However,
before assuming that these are the final velocities we need to 
verify that the newly born compact object did not collide with the companion.
This may happen if two conditions are satisfied: the stars are initially on the ascending
branch of the orbit, and the distance of closest approach $r_{min}=p/(1+\epsilon)$
is smaller than the companion radius.

{\em Single Stars.} For single stars (either initially single or originating
from disrupted binaries), after a SN explosion we simply calculate the
remnant mass, add the natal kick velocity to the spatial speed (non-zero for 
disrupted binaries but zero for primordial single stars) of the object and follow the 
remnant until the end of the calculation.

\subsection{Models}

\subsubsection{Standard Model}

Our standard model (Model~A) corresponds to the one in Paper~I (also called Model~A,
with a few minor differences mentioned in \S,2.1. Here we just reiterate the basic  
standard model parameters. 
We evolve both primordial binaries and single stars. 
Initial stellar masses for single stars and binary primaries are drawn from
a three-component, broken power-law IMF with exponent $\alpha_3=-2.35$ for
massive stars, and a flat mass ratio distribution is used to generate secondary
masses in binaries. The stars are allowed to reach an initial maximum mass 
$M_{\rm max}=150 \msun$ (see \S,2.4.3 for details). 
Stars are evolved with metallicity $Z=0.001$ and our adopted standard wind mass
loss rates (e.g., Hurley et al.\ 2000).
For the treatment of CE phases we use the standard energy prescription 
(Webbink 1984) with $\alpha
\times \lambda = 1.0$. We adopt $M_{\rm max,NS}=3 \msun$ for the maximum NS mass, and
any compact object with mass above this is classified as a BH. As described above
we use lowered kicks for fall-back BHs and no kicks  for direct BHs.

\subsubsection{Alternative Models}

We also performed a set of calculations for a number of different models
in order to test the influence of our most important assumptions and model parameters
on BH formation. Each alternative model differs from our standard reference model in 
the value of one particular parameter or with a change in one particular assumption. 
All models are described in Table~\ref{models}. Note that we added one model not     
present in Paper~I: in Model~J, we use an alternative prescription for CE phases based
on angular momentum balance with parameter $\gamma=1.5$ (see Belczynski, Bulik 
\& Ruiter 2005 and references therein), as opposed to the standard  
energy balance used in all other models. It has been claimed that this alternative CE
prescription leads to better agreement with the observed properties of WD binaries in
the solar neighborhood (Nelemans \& Tout 2005). We wish to test whether this 
new CE prescription has any effect on our predictions for young BH populations.
For some of alternative models (Models~B, D, I and J) the evolution of single 
stars is not affected and we use the single star population from the standard
model. However, we calculate the separate single star populations for models
in which the single star evolution is changed with a given parameter
(Models~C1, C2, E, F, G1, G2 and H).

\subsubsection{Initial Conditions and Mass Calibration}

In addition to a brief description of each model, Table~\ref{models} also gives
the total initial mass in single and binary stars. All stars are assumed to form 
in an instantaneous burst of star formation. Single stars and binary components 
are assumed to form from the hydrogen burning limit ($0.08 \msun$) up to the 
maximum mass $M_{\rm max}$ characterizing a given system. Masses of single stars 
and binary primaries (more massive components) are drawn from the three-component, 
power-law IMF of Kroupa, Tout, \& Gilmore (1993) (see also Kroupa \& Weidner 2003) 
with slope $\alpha_1=-1.3$ within the initial mass range $0.08-0.5\msun$, 
$\alpha_2=-2.2$ for stars within $0.5-1.0 \msun$, and $\alpha_3=-2.35$ within 
$1.0\,M_\odot - M_{\rm max}$. The binary secondary masses are generated from an 
assumed flat mass ratio distribution ($q=M_{\rm a}/M_{\rm b}$; $M_{\rm a},
\ M_{\rm b}$ denoting the mass of the primary and secondary, respectively). The 
mass ratio is drawn from the interval $q_{\rm min}$ to 1, where 
$q_{\rm min}=0.08\msun/M_{\rm a}$, ensuring that the mass of the secondary does not 
fall below the hydrogen burning limit. The only exception is model~B, in which both 
the primary and  the secondary masses are sampled independently from the assumed 
IMF (i.e., the component masses are not correlated). This IMF is easily integrated 
to find the total mass contained in single and binary stars for any adopted 
$\alpha_1, \alpha_2,  \alpha_3$ values. The particular choice of low-mass slope of 
the IMF ($\alpha_1, \alpha_2$) does not change our results, since low-mass stars do 
not contribute to the BH populations. However, as most of the initial stellar mass 
is contained in low-mass stars, a small change in the IMF slope at the low-mass end 
can significantly change the mass normalization.

In our simulations, we do not evolve all the single stars and binaries described 
above since the low-mass stars cannot form BHs. Out of the total population 
described above we evolve only the single stars with masses higher than $4 \msun$ 
and the binaries with primaries more massive than $4 \msun$ (no constraint is placed 
on the mass of the secondary, except that it must be above $0.08 \msun$). 
All models were calculated with $10^6$ massive primordial binaries. We also evolved 
$2 \times 10^5$ massive single stars but then scaled up our results to represent 
$10^6$ single stars\footnotetext{Statistics is much better for single stars than 
binaries; and even with only $2 \times 10^5$ single stars we obtain usually thousands, 
and minimum several hundred, BHs. For example see Tables~\ref{std01}-\ref{std04}}. 
The mass evolved in single stars and binaries was then calculated and, by 
extrapolation of the IMF (down to hydrogen burning limit), the total initial 
cluster mass was determined for each model simulation.

In the discussion of our results we assume an initial (primordial) binary fraction of
$f_{\rm bin}=50\%$, unless stated otherwise (i.e., tables and figures usually assume 
equal numbers of single stars and binaries initially, with 2/3 of stars in binaries). 
However, our results can easily be generalized to other primordial binary fractions 
$f_{\rm{bin}}$ by simply weighing differently the numbers obtained for single stars 
and for binaries.

Our assumed distribution of initial binary separations follows Abt (1983). 
Specifically, we take a flat distribution in $\log a$, so that the probability 
density $\Gamma(a) \propto {1\over a}$. This is applied between a minimum value, such 
that the primary's initial radius (on the zero-age main sequence) is half the radius 
of its Roche lobe, and a maximum value of $10^5 \, {\rm R}_\odot$. We also adopt a 
standard thermal eccentricity distribution for initial binaries, $\Xi(e) = 2e$, in the 
range $e = 0-1$ (e.g., Heggie 1975; Duquennoy \& Mayor 1991).

\subsubsection{Cluster Properties}

The only cluster parameter that enters directly in our simulations is the
escape speed $V_{\rm esc}$ from the cluster core. All single and binary
BHs are assumed immediately
ejected from the cluster if they acquire a speed exceeding $V_{\rm esc}$.
We do not take into account ejections from the cluster halo (where the
escape speed would be lower) as all BHs and their progenitors are expected
to be concentrated near the cluster center.

In Tables~\ref{std01},~\ref{std02},~\ref{std03},~\ref{std04} we present 
results of simulations for our standard model corresponding to four 
different values of the escape speed:
$V_{\rm esc} = 10,~50,~100,~300 \kms$. For any assumed cluster model
the escape speed can be related to the total mass $M_{\rm cl}$ and 
half-mass radius $R_{\rm h}$:
\begin{equation}
V_{\rm esc} = f_{\rm cls} \,\left(\frac{M_{\rm cl}}{10^6\msun}\right)^{1/2}
               \,\left(\frac{R_{\rm h}}{1\,{\rm pc}}\right)^{-1/2}.
\end{equation} 
For example, for a simple Plummer sphere we have $f_{\rm cls} = 106\kms$,
while for King models with
dimensionless central potentials $W_0=3,~5,~7,~9$ and~11, the values are
$f_{\rm cls} = 105.2$, 108.5, 119.3, 157.7, and $184.0\kms$, respectively.
For our four considered values of the escape speed, $V_{\rm esc} = 10$, 50, 
100, and $300\kms$, in a $W_0=3$ King model with $R_{\rm h}=1\,$pc (typical 
for a variety of star clusters), the corresponding cluster masses are 
$M_{\rm cl}= 0.009$, 0.226, 0.904, and $8.132\ \times 10^6 \msun$, 
respectively.

In each table we present the properties of BH populations at five different 
cluster ages: 8.7, 11.0, 15.8, 41.7 and $103.8\,$Myr. These correspond to MS
 turnoff masses 
of 25, 20, 15, 8 and $5\,M_{\odot}$, respectively. 
The tables include information on both the BHs retained in the clusters 
(with velocities $< V_{\rm esc}$) and those ejected from clusters.

\section{RESULTS}

\subsection{Standard Reference Model}

\subsubsection{Black Hole Spatial Velocities}

In Figure~\ref{t1vel} we show distributions of spatial velocities 
for {\em all\/} single and binary BHs shortly after the initial 
starburst (at $8.7\,$Myr). The distribution shows a rather broad 
peak around $\sim 30-300\kms$, but also includes a large fraction 
($\sim$ 2/3) of BHs formed with no kick. The peak originates from a 
mixture of low-velocity binary BHs and high-velocity single BHs. 
The no-kick single and binary BHs originate from the most 
massive stars, which have formed BHs silently and without a kick. All the 
no-kick systems (with zero velocity assumed) were placed on the extreme left side 
of all distributions in 
Figure~\ref{t1vel} to show their contribution in relation to other non-zero 
velocity systems (the bin area is chosen so as to represent their actual 
number, although the placement of the bin along the velocity axis is arbitrary).
Binary stars hosting BHs survive only if the natal kicks they received 
were relatively small, since high-magnitude kicks tend to disrupt the
systems. We see (middle panel of Fig.~\ref{t1vel}) that most BH 
binaries have spatial velocities around $50\kms$, which
originate from the low-velocity side of the bimodal Arzoumanian et al.\ 
(2002) distribution. 
Single BHs originating from single stars follow closely the bimodal 
distribution of natal kicks, but the final BH velocities are  
slightly lower because of fall-back and direct BH
formation (see \S2.3). The low- and high-velocity single BHs have 
speeds around $50\kms$ and $250\kms$, respectively.
Single BHs originating from binary disruptions gain high speeds ($\sim 
100 - 400\kms$), since binaries are disrupted when
a high-magnitude kick occurs. Finally, the single BHs formed through binary 
mergers have the lowest (nonzero) velocities ($\sim 10 - 100\kms$), since they are the
most massive BHs and therefore most affected by fall-back. 

In Figure~\ref{t5vel} we show the velocity distributions at a later time  
($103.8\,$Myr) when essentially all BHs have formed, and no more SNe 
explosions are expected, so the velocity distribution is no longer
evolving (the MS turnoff mass for that time is down to 
$5\msun$). The velocities have now shifted to somewhat 
higher values (with a single peak at $\simeq 200\kms$ for non-zero velocity BHs), 
while the relative contribution of no-kick systems drops to 
around 1/3.
At this later time the population is more dominated by 
single BHs. Most of the non-zero velocity single BHs come from binary 
disruptions (see middle panel of Fig.~\ref{t5vel}) and therefore they have 
received larger kicks, shifting the overall distribution toward slightly 
higher velocities.
Also, at later times, lower-mass BH progenitors go through SN explosions, 
and they receive on average larger kicks (since for lower masses there is 
less fall back).

We note that most of the non-zero-velocity BHs gain speeds of $50-200\kms$ 
in SNe explosions. Depending on the properties of a given 
cluster they may be ejected or retained, and will either populate the field 
or undergo subsequent dynamical evolution in the cluster.
We now discuss separately the properties of the retained and ejected BH
populations.

\subsubsection{Properties of Retained (Cluster) BH Populations}

Retained BHs in clusters could be found either in binaries or as single objects.
Binary BHs are found with different types of companion stars, while single
BHs may have formed through various channels, which we also list in 
Tables~\ref{std01} -- ~\ref{std04}.  
Shortly after the starburst the most frequent BH companions are massive MS stars, 
but, as the population evolves, these massive MS companions finish their
lives and form additional BHs. Double BH--BH systems begin to 
dominate the binary BH population after about $15\,$Myr.
At later times less massive stars evolve off the MS and start contributing to 
the sub-population of BHs with evolved companions (CHeB stars being the dominant
companion type, with a relatively long lifetime in that phase) or other remnants
as companions (WDs and NSs). Once the majority of stars massive enough to make 
BHs end their lives (around $10-15\,$Myr) we observe a general 
decrease in the total number of BHs in binaries. The number of 
BHs in binaries is depleted through the disruptive effects of SNe and binary
mergers (e.g., during CE phases). Both processes enhance the single 
BH population. This single population is dominated by the BHs formed from
primordial single stars (assuming $f_{\rm bin}=50\%$). 
The formation along this single star channel stops early on 
when all single massive stars have finished evolution and formed BHs (at $\simeq 
10-15\,$Myr). In contrast, the contribution 
of single BHs from binary disruptions and mergers is increasing with time, 
but eventually it also saturates (at $\simeq 50-100\,$Myr), since there are fewer
potential BH progenitor binaries as the massive stars die off.  
In general, the single BHs are much more numerous in young cluster
environments than binary BHs. At early times ($\simeq 10\,$Myr) they 
dominate by a factor $2-4$, but later the ratio of single to  
binary BHs increases to almost 10 (after $\sim 100\,$Myr), as
many binaries merge or are disrupted (adding to the single population).

\subsubsection{Properties of Ejected (Field) BH Populations}

Tables~\ref{std01} -- \ref{std04} show also the properties of BHs 
 ejected from their parent clusters, assuming different escape speeds.
Significant fractions ($\ga 0.4$) of single and binary BHs are likely to be ejected 
from any cluster with escape speed $V_{\rm esc} \la 100\kms$. 
In general, single BHs are more prone to ejection since they gain larger 
speeds in SNe explosions (compared to heavier binaries). 
Early on the number of ejected BHs increases with time as new BHs of lower 
mass (and hence receiving larger kicks) are being formed. At later times 
(after $\simeq 15\,$Myr), the number of fast BHs remains basically unchanged. 
Ejected binaries consist mostly of BH--MS and BH--BH pairs in comparable 
numbers. Rare BH--NS binaries are ejected more easily 
than other types since they experience two kicks. 
Single ejected BHs consist mostly of BHs originating from single stars 
which have received large kicks and from the components of a disrupted 
binary (the involved kicks were rather large to allow for disruption). 

\subsubsection{Dependence on Cluster Escape Velocity and Initial Binary
Fraction}

In Table~\ref{frac01} we list fractions of retained BHs at $103.8\,$Myr 
after the starburst. The results are presented for initial 
cluster binary fractions of $f_{\rm bin}=0,\ 50,\ 100\%$, and can be 
linearly interpolated for the desired $f_{\rm bin}$.  
For our standard model the results are shown for the four considered escape 
velocities. For an initial cluster binary fraction $f_{\rm bin}=50\%$ we find 
that the retained fraction can vary from $\sim 0.4$ for low escape 
velocities ($V_{\rm esc} = 10\kms$) to $\sim 0.9$ for high 
velocities ($V_{\rm esc} = 300\kms$). For escape 
velocities typical of globular clusters or super star clusters
($V_{\rm esc} \sim 50 \kms$), retained and ejected 
fractions are about equal. 
The retained fractions for various types of systems are plotted
as a function of $V_{\rm esc}$ in Figure~\ref{retained.all}. All curves 
are normalized to total number of BHs, both single and binaries. 

Results are listed in Table~\ref{frac01} for different binary 
fractions. In particular, these can be used to study the limiting
cases of pure binary populations ($f_{\rm bin}=100\%$) and pure 
single star populations ($f_{\rm bin}=0\%$). 
Note that even an initial population with all massive stars in binaries
will form many single BHs
through binary disruptions and binary mergers. 
We also note the decrease of the retained fraction with increasing initial 
binary fraction. Clusters containing more binaries tend to lose relatively 
more BHs through binary disruptions in SNe compared to single star 
populations.

\subsubsection{Orbital Periods of Black Hole Binaries} 

Figure~\ref{A.50.per} presents the period distribution of BH binaries for our 
standard model (for the characteristic escape velocity $V_{\rm esc} = 50\kms$). 
We show separately populations retained and ejected from a 
cluster. The distributions for different values of the escape velocity are 
similar. 

In Paper~I we obtained a double-peaked period distribution  
for BHs in field populations: tighter binaries were found around $P_{\rm orb} \sim 10\,$d while 
wider systems peaked around $P_{\rm orb} \sim 10^5\,$d. The shape of this
distribution comes from 
the property that tighter BH progenitor systems experienced at least one RLOF/CE
episode leading to orbital decay, while wider systems never interacted
and stayed close to their initial periods. The two peaks are clearly
separated with a demarcation period $P_{\rm s} \sim 10^3\,$d. 

It is easily seen here that slow and fast BH populations add up to the original
double-peaked distribution of Paper~I. Only the shortest-period and hence 
most tightly bound systems ($P_{\rm orb} < P_{\rm s}$) survive SN 
explosions and they form a population of fast, short-period BH binaries (see 
bottom panel of Fig.~\ref{A.50.per}). In contrast, systems retained in
clusters have again a double-peaked orbital period distribution. The slowest 
systems have rather large periods ($P_{\rm orb} \sim 10^5\,$d) and they 
will likely get disrupted through dynamical interactions in the dense
cluster core. The short-period cluster binaries ($P_{\rm orb} \sim 10-100\,$d) 
are much less numerous, since most of the short-period systems gained high 
post-SN velocities and contributed to the ejected population. Compared to 
Paper~I we note that the inclusion of ejections further depletes the cluster 
{\em hard\/} binary BH population. Only about 1/3 of systems are found with 
periods below $P_{\rm s}$, half of which are retained within a cluster 
with $V_{\rm esc} = 50\kms$. For a cluster with $V_{\rm esc} = 100\kms$ 
about 80\% of the short-period systems are retained.

\subsubsection{Black Hole Masses}

Black hole mass distributions are presented in Figures~\ref{A.50.mass.all}, 
~\ref{A.50.mass.sin} for $V_{\rm esc} = 50\kms$. With few 
exceptions the models for different escape velocity values are very 
similar. The retained and ejected populations are shown in separate panels. 

The retained populations of BHs shown in the top panel of Figure~\ref{A.50.mass.all} 
have a characteristic triple-peaked mass distribution: a first peak 
at $M_{\rm BH} \sim 6-8 
\msun$, a second one at $M_{\rm BH} \sim 10-16 \msun$, and third at $M_{\rm BH} \sim 
22-26 \msun$; beyond this it steeply falls off with increasing mass.
 
The shape of the distribution is determined by the combination of IMF and
initial-to-final mass relation for single BHs (presented and discussed in
detail in Paper~I): the most massive stars ($\geq 50\msun$) form BHs with masses
in the range $\sim 10-16\msun$; stars within an initial mass range $25-35\msun$ 
form BHs of mass $\sim 25\msun$; stars of initially $40-50\msun$ 
tend to form $7\msun$ BHs. Both single and binary BHs contribute
significantly to the second and third peaks. However, only single stars 
are responsible for a first narrow peak corresponding to a pile up of 
BHs in the initial-to-final mass relation around $6-8\msun$. 
This characteristic feature is a result of a very sharp transition in
single star evolution, from H-rich to naked helium stars, which is caused
by wind mass loss and the more effective envelope removal for single stars
above a certain initial mass.
In binary stars, removal of the envelope can happen not only through
stellar winds but also through RLOF, and so it is allowed for the
entire mass range and the first peak is washed out.

The ejected populations, shown in the bottom panel of Figure~\ref{A.50.mass.all},
are dominated by single BHs (due to their high average speeds) with masses 
$\sim 3-30\msun$. The distributions have one sharp peak at $M_{\rm BH} \sim 6-8\msun$, 
corresponding to the first low-mass peak in the distribution for retained 
populations. The high-mass BHs are very rare in the ejected 
populations since the kick magnitudes decrease with increasing BH mass (because of 
significant fall-back or direct BH formation at the high-mass end).

BHs in binary systems reach a maximum mass of about $30\msun$ for 
both cluster and ejected populations.  
Most single BHs have masses below $30\msun$. However, the
tail of the single BH mass distribution extends to $\sim 50\msun$ for ejected 
populations and to about $80\msun$ for cluster populations. This is
shown in Figure~\ref{A.50.mass.sin} (note a change of vertical scale as 
compared to Figure~\ref{A.50.mass.all}).  
The highest-mass BHs are always retained in the clusters and they are
formed through binary mergers. These mergers are the result of 
early CE evolution of massive binaries. The most common merger types are 
MS--MS, HG--MS and BH--HG mergers. During mergers involving HG stars we assume that 
the envelope of the HG star is lost, while the BH/MS star and the compact 
core of the HG star merge to form a new, more massive object. The merger product is 
then evolved and it may eventually form a single BH.

Even with significant mass loss through stellar winds and during the 
merger process, a small fraction of BHs reach very high masses, up to  
about $80\msun$. With a less conservative assumption, allowing some fraction of 
the HG star envelope to be accreted onto the companion in a merger, 
the maximum BH mass could then reach even higher values $\ga 100\msun$. 
In Figure~\ref{A.50.fullmass.sin} we show the results of a calculation with the 
merger product's mass always assumed equal to the total binary mass. 

Although the amount of mass loss in a merger is rather uncertain, the 
two models above (with and without mass loss) indicate that binary 
star evolution could lead to the formation of single $\sim 100\msun$ BHs.
These most massive BHs form very early in the evolution of a cluster (first $\sim 
5-10\,$Myr) since they originate from the most massive and rapidly evolving stars. 
These BHs are retained in clusters (direct/silent BH formation with no
associated natal kick) and they may act as potential seeds for building up
intermediate-mass BHs through dynamical interactions during the subsequent cluster
evolution (Miller \& Hamilton 2002; O'Leary et al.\ 2005).

\subsection{Parameter Study}

\subsubsection{Black Hole Spatial Velocities}

For most alternative models the velocity distributions are similar to those found
in the reference model (see Fig.~\ref{t1vel} and ~\ref{t5vel}). These
distributions are generally characterized by the same wide, high-velocity peak (tens to
hundreds of $\kms$) and a rather large population of zero-kick BHs. 
In particular, for models~B, F, G2, J and~H, the distributions are almost
identical to those of the reference model at all times.
For models~D, G1, and~I the distributions show slight differences. With
lowered CE efficiency (model~D) it is found that there are fewer fast binary
BHs, and most surviving binaries do not gain higher velocities at early
times. Basically, many tight binaries that survived SN explosions in
the reference model have now merged in a first CE phase, even before the 
first SN explosion occurred.  
In model G1, in which we consider only primordial stars up to $M_{\rm max}=
50\,\msun$, the population of massive BHs formed through direct collapse
(with no kick) is significantly reduced. This results in
a velocity distribution similar to that of the reference model for non-zero 
velocity systems, but with a much lower number of zero-kick BHs.  
The model~I distribution is slightly different, especially at early
times when most BHs form with no kick, since in this model we
consider only the most massive BHs formed mainly through direct collapse.  
 
A few models show more significant differences. Different metallicities lead to  changes 
in BH velocities, especially at early times. For very low metallicity (model 
C1) almost all BHs are formed with no kick, while for high, 
solar-like metallicity (model C2) most BHs have non-zero velocities in a wide range 
($\sim 10-1000 \kms$). Metallicity strongly affects the wind mass loss rates,
which are most important for the evolution of the most massive stars (i.e., at early times). In
particular, for low-$Z$ values, the wind mass loss rates are smaller (hence more 
high-mass pre-SN
stars and direct collapses), while for high $Z$ the winds are very effective in
removing mass from BH progenitors (hence smaller mass pre-SN stars, and more
fall-back BH formation).   
The most significant difference is found in model~E, where
we allow for full BH kicks. All BHs are formed with
rather high ($\geq 100 \kms$) velocities. The distribution, shown in
Figures~\ref{t1velE} and ~\ref{t5velE}, is double-peaked both for early and 
late times. The single stars dominate the population, forming the low- 
($\sim 100 \kms$) and high-velocity component ($\sim 500 \kms$), a direct
result of the adopted bimodal natal kick velocity distribution. Binary stars are
found at lower velocities ($\sim 100 \kms$) but they are only a minor
contributor to the overall BH population since most of them are now disrupted 
at the first SN explosion.

\subsubsection{Properties of Retained (Cluster) BH Populations}

In Table~\ref{parm01} we present the properties of cluster BH populations $11\,$Myr 
after the starburst. 
Results for the various models may be easily compared with our reference model. 
 
Binary BHs for different model assumptions are still in general dominated by 
BH--MS and BH--BH binaries. These systems appear in comparable numbers in most models. 
Only for models~B and~E do we find a smaller contribution of BH--BH binaries 
($\sim$ 5\% and almost zero for models~E and~B, respectively). 
In model~B the independent choice of masses produces systems with extreme mass ratios, 
so that massive primordial binaries with two BH progenitors are very rare. 
Obviously for model E, in which the two BHs receive full kicks, the BH--BH binary
formation is strongly suppressed by binary disruptions. 
The highest number of binaries containing BHs is found in our model with the 
lowest tested metallicity (C1). For low metallicities BHs form preferentially with
high masses (low wind mass loss rates) through direct collapse with no kick. 
In contrast model~E, assuming full BH kicks, results in the lowest number of 
BH binaries. 
Many models (D, G2, H, I, J) result in very similar contents to our reference model. 
It is worth noting in particular that the CE treatment (either lowered 
efficiency in model~D, or different prescription in model~J) does not appear to
play a significant role in determining cluster initial binary BH populations.   

For all models the single BHs dominate the population even at very early times 
(as early as $11\,$Myr). Single BHs originate predominantly from primordial single 
stars, with smaller 
contributions from disrupted binaries and binary mergers. The basic general trends 
seen in our reference model are preserved in other models. Also most models (B, C1, D, G2, H, 
I, J) form similar numbers of single BHs as our reference model.    
It is found, as in the binary populations, that the highest number of
 single BHs is seen in 
our model with lowest metallicity (C1), while the model with full BH kicks (E) generates the 
lowest number of single BHs retained in a cluster. 

At $103.5\,$Myr (see Table~\ref{parm02}), when no more BHs are being formed,
single BHs strongly dominate (by about an order of magnitude) over binary BHs.
Single BHs still originate mostly from primordial single stars, but there is an 
increased contribution from binary mergers and disruptions.
The binary population remains dominated by BH--BH and BH--MS systems in most models, 
but with an increased contribution from other evolved systems (BH--WD and BH--NS) 
compared to earlier times. Note that only in model~E does the number of systems 
other than BH--BH and BH--MS end up dominating the binary population.

\subsubsection{Properties of Ejected (Field) BH Populations}

In Tables~\ref{parm01} and ~\ref{parm02} we also characterize the populations of ejected 
(field) BHs for various models. Results for both times are 
comparable for binary BHs, but with significantly more single BHs being ejected at 
later times. 

The BH--MS and BH--BH binaries, which dominate the total populations, are also 
found to be most effectively ejected from clusters. However, BH--NS systems, receiving 
two natal kicks, are also found to be easily ejected. Indeed, in many models 
(C1, C2, E, F, G1, G2, H, J), they constitute a significant fraction of ejected systems. 
Contrary to our intuitive expectation, evolution with the full BH kicks (model~E) does 
not generate a particularly large population of fast BH binaries. In fact, the ejected population 
is smaller than in the reference model. Higher kicks are much more 
effective in binary disruption than in binary ejection.

The numbers of fast single BHs are comparable in most models (A, B, D, G1, G2, H, J),
with the ejected populations usually consisting equally of BHs coming from 
binary disruptions and primordial single stars, with a smaller contribution from 
merger BHs. For models with massive BHs (C1 and I) which receive small kicks there 
are fewer single BHs in the ejected population (by a factor $\sim 2$). On the other 
hand, for the model with full BH kicks (E), the ejected single BH population is larger 
(by a factor of $\sim 3$) compared to the reference model.

\subsubsection{Dependence on Cluster Escape Velocity and Initial Binary Fraction}

Retained fractions for different evolutionary models follow in general the
same trends as in our reference model, i.e., retained fractions decrease with 
increasing initial binary fraction. The exception to that trend is for models with 
full BH kicks (E), increased metallicity (C2) or uncorrelated binary
component masses (B).
Also, independent of the escape velocity, it is found that at least $\sim 40\%$ of BHs 
are retained simply because of no-kick BHs (for an initial binary fraction of 50\%), 
with the obvious exception of 
the model with full BH kicks (E). In particular, for models~C1, D, F, G1, G2, H, I, 
and~J, the dependence of the retained fraction on $V_{\rm esc}$ is very similar to that
seen in the reference model (see Fig.~\ref{retained.all}).

In model~B, th secondary mass is on average very small compared to the BH mass
(due to our choice of initial conditions for this model). Therefore, BHs in
binary systems gain similar velocities (almost unaffected by their companions) 
as single BHs, and this leads to almost constant fraction of retained
systems ($\sim 0.64$) independent of the initial binarity of the cluster.    
In models C2 and E the fraction of retained systems may be as small as 0.4 
and 0, respectively. In our model with high metallicity (C2), as discussed
above (\S\,3.2.1), high wind mass loss rates lead to higher BH kicks and
hence smaller retained fractions. The most dramatic change is observed for
model~E, with full BH kicks. The retained fractions for this model
are shown in Figure~\ref{retained.allE}. Here, we also
normalize all curves to the total number of BHs (single and in binaries).
The retained fraction increases from 0 to $\sim 0.9$, approximately proportional 
to the escape velocity, with no apparent flattening up to $V_{\rm esc} \sim 1000 
\kms$ as a result of the high speeds BHs receive at formation. The total
retained fraction does not reach unity, since there is still a small number of 
BHs with velocities over 1000 km s$^{-1}$.
Larger BH kicks (switching from standard lowered kicks to full kicks) decrease the 
retained fraction from 0.6 to 0.2 for $V_{\rm esc} \simeq 50 \kms$ and $f_{\rm bin}50\%$. 

A summary of retained and ejected fractions for different initial cluster binary 
fractions is presented in Table~\ref{frac01} for $V_{\rm esc}=50\kms$. In particular, 
we show results for pure single star populations ($f_{\rm bin}=0\%$) and for all binaries
($f_{\rm bin}=100\%$). Note that single star populations will obviously form 
only single BHs, while the binary-dominated clusters will form both BH binaries and single 
BHs (through disruptions and mergers)\footnote{The number of single BHs formed out of binary 
systems may be inferred by comparing the numbers of binary BHs with the single BHs 
listed under ``binary disruption'' and ``binary mergers'' 
in Tables~\ref{std01}--\ref{std04}, ~\ref{parm01} and ~\ref{parm02}}.

For the standard $f_{\rm bin}=50\%$ it is found that the retained fraction of BHs varies
from 0.4 -- 0.7 across almost all models. The only exception is model~E with full BH kicks for 
which the retained fraction is only 0.2. For more realistic and higher 
initial cluster binary fractions ($f_{\rm bin}=75-100\%$; see Ivanova et al.\ 2005) 
the retained BH fraction is 
found in an even narrower range 0.4 -- 0.6 (again with the exception of model E).
Therefore, despite the number of model uncertainties, the initial BH cluster populations, 
as far as the numbers are concerned, are well constrained theoretically. The issue of BH 
kicks is not resolved yet, but both observational work (e.g., Mirabel \& Rodrigues 2003) and 
theoretical studies (e.g., Willems et al.\ 2005) are in progress.

\subsubsection{Orbital Periods of Black Hole Binaries}

In Figures~\ref{mod.per1} and ~\ref{mod.per2} we show the dependence of the period 
distributions of BH binaries on model assumptions.
In general, the period distribution remains bimodal in most of the models (B,
C1, C2, D, F, G1, G2, H)  for retained BH binaries, while only short-period 
binaries tend to be ejected from clusters, as explained in \S\,3.1.5.
Most of the retained binaries are formed with rather large orbital periods (with 
the exception of model~E, see below) and they will be prone to dynamical
disruption in dense cluster environments. 

The major deviations from the reference model are found for model~E, with full 
BH kicks. The retained population is rather small as compared to the other 
models and consists mostly of short-period binaries, since all of the wide BH 
systems were disrupted by SN natal kicks. The majority of short-period binaries 
which survived gained significant velocities ($\geq 50 \kms$; see 
Fig.~\ref{t5velE}) and the ejected population is the most numerous in this model.

In several other models we find smaller variations from the reference period 
distribution. Models with different CE efficiency and treatment (D and J), in
which most close binaries merge, have very small numbers of short-period binaries.
Also, the model in which we consider only the most massive BHs
(descendants of wide primordial binaries) is characterized by a smaller 
short-period binary population.

\subsubsection{Black Hole Masses}

In Figures~\ref{mod.mas.all1} and~\ref{mod.mas.all2} we present the BH mass 
distributions from all the models in our study, for both single and binary BH
populations.

The shape of the distribution for the retained and ejected BH populations is not 
greatly affected by different choices of parameter values, with the exception 
of metallicity and BH kicks (see Fig.~\ref{mod.mas.all1}). This is easily 
understood, as the highest-mass BHs are formed only at low metallicity 
(Models~C1 and~A) and the lightest BHs are formed at high metallicity 
(Model~C2). For full BH kicks (model E) the majority of BHs gain high speeds, 
and the mass distribution for the ejected population is similar to the 
combination of ejected and retained populations in the reference model. 

Most of the BHs do not exceed $\sim 25 \msun$. However, a small fraction of  
single BHs in many models reach very high masses around $80\,\msun$. 
Figures~\ref{mod.mas.sin1} and~\ref{mod.mas.sin2} show the mass
distributions of single BH subpopulations. 
In all the models with high-mass BHs the most massive BHs are formed through
binary mergers. In all the calculations presented here we have assumed mass
loss during the merger process if an evolved star was involved (as discussed in
\S\,3.1.6). The highest maximum BH masses are found in the lowest metallicity 
environments (models A and C1), in larger systems (with high $M_{\rm max}$,
model G2), and for binaries formed with full BH kicks (model E),
quite independent of other evolutionary parameters.
We find $\sim 10-100$ BHs with masses over $60 \msun$ in models C1, C2, E, G2  
and J (and fewer in other models), for a total starburst mass of $\sim 10^8 
\msun$ (see Table~\ref{models}).
The highest mass BHs are retained in clusters with the exception of the
model incorporating full BH kicks, in which they are found both with high
and low speeds.
Only in a few models, with uncorrelated initial binary component masses (B), 
low CE efficiency (D), or low $M_{\rm max}$ (G1), does the maximum BH mass stay 
below $\sim 60 \msun$. And in particular, in the model~B, the maximum BH
mass stay below $\sim 30 \msun$. This is due to the fact, that in this model 
BHs are accompanied by relatively low mass companions and therefore there is no
mass reservoir to increase substantially initial (formation) BH mass.

\section{SUMMARY AND DISCUSSION}

Using the population synthesis code {\tt StarTrack} we have studied the formation 
of single and binary BHs in young star clusters.
Our study continues and improves on the initial work described in Paper~I
by taking explicitly into account the likely ejections of 
BHs and their 
progenitors from star clusters because of natal kicks imparted by 
SNe or recoil following binary disruptions. The results indicate that 
the properties of both retained BHs in clusters and ejected BHs (forming a
field population) depend sensitively on the depth of the cluster potential. 
For example we find that most BHs ejected from binaries are also ejected from 
clusters with central escape speeds $V_{\rm esc}\la 100\,{\rm km}\,{\rm s}^{-1}$, 
while most BHs remaining in binaries are retained by clusters with
$V_{\rm esc}\ga 50\,{\rm km}\,{\rm s}^{-1}$. Also, approximately half of the 
single BHs originating from the  primordial single star population are ejected 
from clusters with $V_{\rm esc}\la 50\,{\rm km}\,{\rm s}^{-1}$.
The overall BH retention fraction increases gradually from $\sim 0.4$ to 0.7 
as the cluster escape speed increases from $\sim 10$ to $100 \kms$ (Fig.~4).
Tables~2--5 give the numbers
of BHs in different kinds of systems, both retained in and ejected from clusters
with different escape speeds. Their main properties are illustrated in Figures~5--8.
Single BH masses can become as large as $\sim 100\,M_\odot$ (as a consequence of
massive binary mergers, especially if mass loss during mergers is small). These 
``intermediate-mass'' BHs are almost always retained in clusters. If they were to 
acquire a new binary companion through dynamical interactions in the dense
cluster environment, they could become ULXs. However, it was recently
demonstrated that although massive BHs easily acquire binary companions, it is 
rather unlikely to find such a binary at high ultraluminous X-ray luminosity 
(Blecha et al. 2006). 

BH--BH binaries (rather than double NSs), are probably
the most promising GW sources for detection by ground-based
interferometers (Lipunov, Postnov, \& Prokhorov 1997; Bulik \& Belczynski
2003). Merging BH--BH systems therefore are important sources for present
projects to detect astrophysical GW sources (e.g., GEO, LIGO, VIRGO).
The properties of BH--BH binaries in
much larger stellar systems with continuous star formation (e.g.,
disk galaxies) were studied  extensively by Bulik \& Belczynski
(2003; Bulik, Belczynski \& Rudak 2004a; Bulik, Gondek-Rosinska \& 
Belczynski 2004b). We find that the properties of BH--BH binaries in 
starbursts are not too different from those found in previous studies. 
Most BH--BH systems are characterized by rather equal masses, with a mass ratio 
distribution peaking at $q \simeq 0.8-1.0$ (cf.\ the Pop~II models of Bulik et 
al.\ 2004b). For most models only a small fraction (a few per cent; e.g., 
5\% for Model~A) of the BH--BH systems are tight enough to merge within a 
Hubble time and produce observable GW signals. For models which tend to produce 
tighter BH--BH binaries (D and E), the fraction can be significantly higher 
($\sim$ 10--40 \%).
However, in Model~E there are almost no BH--BH binaries, and the higher
fraction of coalescing systems does not mean a higher BH--BH merger rate.
Models C1 and D are the most efficient in producing merging BH--BH binaries: 
2035 and 1370, respectively (for a total starburst mass of $\sim 10^8
\msun$; Table~\ref{models}), while for most of the other models (including
the reference model) we find $\sim 300-600$ merging BH--BH systems.

\acknowledgements 
This work was supported in part by KBN grants PBZ-KBN-054/P03/2001 and 
1 P03D 022 28 at the Copernicus Center, Poland, and by NSF Grant PHY-0245028 and 
NASA Grants NAG5-12044 and NAG5-13236 at Northwestern University. 
For hospitality and support, AS thanks the Astronomy Department at New Mexico 
State University and the Theoretical Astrophysics Group at Northwestern University; 
KB thanks the Aspen Center for Physics; and FAR thanks the 
Aspen Center for Physics and the Center for Gravitational Wave Physics at Penn State University.

\clearpage

\begin{deluxetable}{llcc}
\tablewidth{390pt}
\tablecaption{Population Synthesis Model Assumptions}
\tablehead{ 
      &                              & Mass [$M_\odot$] in  & Mass [$M_\odot$] in \\
Model & Description\tablenotemark{a} & Single Stars & Binaries }
\startdata
A      & standard model described in \S\,2.4    & $3.79 \times 10^{7}$ & $5.95 \times 10^{7}$ \\
B      & uncorrelated binary component masses   & $3.79 \times 10^{7}$ & $7.59 \times 10^{7}$ \\
C1-2   & metallicity $Z = 0.0001, 0.02$         & $3.79 \times 10^{7}$ & $5.95 \times 10^{7}$ \\
D      & standard CE: $\alpha_{\rm CE} \times \lambda = 0.1$ & $3.79 \times 10^{7}$ & $5.95 \times 10^{7}$ \\
E      & full kicks for BHs                     & $3.79 \times 10^{7}$ & $5.95 \times 10^{7}$ \\
F      & steeper IMF: $\alpha_3=-2.7$         & $5.97 \times 10^{7}$ & $9.49 \times 10^{7}$ \\ 
G1     & lower maximum mass: $M_{\rm max} = 50 M_\odot$  & $3.69 \times 10^{7}$ & $5.80 \times 10^{7}$ \\
G2     & lower maximum mass: $M_{\rm max} = 100 M_\odot$ & $3.75 \times 10^{7}$ & $5.88 \times 10^{7}$ \\
H      & $M_{\rm max,NS}=2 $\,M$_\odot$         & $3.79 \times 10^{7}$ & $5.95 \times 10^{7}$ \\ 
I\tablenotemark{b} & BHs more massive than 10 M$_\odot$ & $3.79 \times 10^{7}$ & $5.95 \times 10^{7}$ \\
J      & alternative CE: $\gamma=1.5$         & $3.79 \times 10^{7}$ & $5.95 \times 10^{7}$ \\ 
\enddata
\label{models}   
\tablenotetext{a}{Details of model assumptions are given in 
\S\,2.4.}
\tablenotetext{b}{Model~I is shown only to give the numbers of 
 BHs (formed in the standard Model~A) with mass greater than 
 $10 M_\odot$.}
\end{deluxetable}
\clearpage

\begin{deluxetable}{lccccc}
\tablewidth{480pt}
\tablecaption{Black Hole Populations: Retained in/Ejected from Cluster\tablenotemark{a}\ with 
              $V_{\rm esc}=10$ km s$^{-1}$ - Standard Model}
\tablehead{ Type\tablenotemark{b} & 8.7 Myr & 11.0 Myr & 15.8 Myr & 41.7 Myr & 103.8 Myr \\
 & $M_{\rm to}=25\msun$ & $M_{\rm to}=20\msun$ & $M_{\rm to}=15\msun$ & 
$M_{\rm to}=8\msun$ & $M_{\rm to}=5\msun$ }
\startdata
BH--MS&                   13113/4897    &  11444/5422    &  8622/4114  & 4696/2238   & 2941/1338  \\   
BH--HG&                   12/2          &           16/5 &    16/3     &  9/6        &10/5       \\
BH--RG&                   0/0           &            0/0 &    0/0      & 1/3         & 3/6      \\
BH--CHeB&                 1193/72       &         945/116&    634/68   & 270/16      & 175/25   \\ 
BH--AGB&                  25/0          &           16/0 &   19/1      & 10/1        & 14/0     \\
BH--He&                   79/142        &        29/107  &   10/90     &  7/281      & 0/388      \\
BH--WD&                   0/0           &            0/0 &    0/9      & 37/396      & 1106/1163  \\
BH--NS&                   0/62          &         0/370  &     2/793   & 11/999      & 16/967     \\
BH--BH&                   8758/1880     &   9180/2252    &  9179/2257  & 9179/2247   & 9179/2226  \\
Total in binaries:&      23180 /7055    &   21630/8272   &  18482/7335 & 14220/6187  & 13444/6118  \\ 
 
&&&&&\\

Single: binary disruption&   2103/26501 &   3164/46773 &  3798/60129  & 3877/65384   & 3878/66190   \\
Single: binary merger&     5507/2236    & 7368/5452    &  13429/8342  & 15522/16234  & 15526/16300        \\  
Single progenitor&         60015/16315  & 63030/43360  &  63030/55855 & 63030/55855  & 63030/55855  \\  
Total single:&             67625/45052  & 73562/95585  & 80257/124326 & 82429/137473 & 82434 /138345  \\ 

\enddata
\label{std01}
\tablenotetext{a}{Retained populations contain all BHs with $V_{\rm bh} < V_{\rm esc}$ while 
ejected populations include BHs with $V_{\rm bh} \geq V_{\rm esc}$}
\tablenotetext{b}{Black holes in binary systems are listed according to their
companion types: MS---main sequence, HG---Hertzsprung Gap, RG---reg giant, CHeB---core 
He burning, AGB---asymptotic giant branch, He---helium star, WD---white dwarf,
NS---neutron star, BH---black hole.
Single black holes formed from components of disrupted binaries are listed
under ``Single: binary disruption.''  
Single black holes formed from binary merger products are under 
``Single: binary merger.'' 
Single black holes that are remnants of single stars are listed under ``Single
progenitor.''
}
\end{deluxetable}

\begin{deluxetable}{lccccc}
\tablewidth{480pt}
\tablecaption{Black Hole Populations: Retained in/Ejected from Cluster\tablenotemark{a}\ with 
              $V_{\rm esc}=50$ km s$^{-1}$ - Standard Model}
\tablehead{ Type & 8.7 Myr & 11.0 Myr & 15.8 Myr & 41.7 Myr & 103.8 Myr \\
 & $M_{\rm to}=25\msun$ & $M_{\rm to}=20\msun$ & $M_{\rm to}=15\msun$ & 
$M_{\rm to}=8\msun$ & $M_{\rm to}=5\msun$ }
\startdata
BH--MS&                            15616/2394 &  14055/2811 & 10357/2379  & 5382/1552   & 3317/962   \\   
BH--HG&                               14/0 &           20/1 &    19/0     & 11/4        & 11/4       \\
BH--RG&                                0/0 &            0/0 &    0/0      & 1/3         & 4/5      \\
BH--CHeB&                           1261/4 &         1057/4 &    697/5    & 284/2       & 187/13   \\ 
BH--AGB&                              25/0 &           16/0 &   20/0      & 11/0        & 14/0     \\
BH--He&                              204/17 &        123/13 &    86/14    & 117/171     & 45/343     \\
BH--WD&                                0/0 &            0/0 &    0/9      &  274/159    & 1682/587   \\
BH--NS&                               1 /61 &        11/359 &    66/729   & 173/837     &  181/802   \\
BH--BH&                             9787/851 &   10410/1022 &  10415/1021 & 10415/1011  & 10415/990  \\
Total in binaries:&               26908/3327 &   25692/4210 &  21660/4157 & 16668/3739  & 15856/3706  \\ 

&&&&&\\
  
Single: binary disruption&  10775/17829 &   18136/31801&  22195/41732 & 22848/46413 & 22900/47168  \\
Single: binary merger&     6804/939     & 9906/2914    &  17059/4722  & 20625/11131 & 20630/11196        \\  
Single progenitor&         65230/11100  & 75465/30925  &  75830/42585 & 75830/42585 & 75830/42585  \\  
Total single:&             82809/29868  &103507/65640  & 115084/89039 &119303/100129&119360/100949  \\ 

\enddata
\label{std02}
\tablenotetext{a}{
Same as Table~\ref{std01}.
}
\end{deluxetable}

\begin{deluxetable}{lccccc}
\tablewidth{480pt}
\tablecaption{Black Hole Populations: Retained in/Ejected from Cluster\tablenotemark{a}\ with 
              $V_{\rm esc}=100$ km s$^{-1}$ - Standard Model}
\tablehead{ Type\tablenotemark{a} & 8.7 Myr & 11.0 Myr & 15.8 Myr & 41.7 Myr & 103.8 Myr \\
 & $M_{\rm to}=25\msun$ & $M_{\rm to}=20\msun$ & $M_{\rm to}=15\msun$ & 
$M_{\rm to}=8\msun$ & $M_{\rm to}=5\msun$ }
\startdata
BH--MS&                            17163/847 & 15857/1009   &  11794/942  & 6178/756   & 3722/557   \\   
BH--HG&                               14/0 &           21/0 &     19/0    & 15/0       & 12 /3     \\
BH--RG&                                0/0 &            0/0 &      0/0    &  2/2       & 7/2      \\
BH--CHeB&                           1265/0 &          1061/0&    702/0    & 286/0      & 190/10   \\ 
BH--AGB&                              25/0 &           16/0 &     20/0    & 11/0       & 14/0     \\
BH--He&                              215/6 &          131/5 &     97/3    & 252/36     & 222/166    \\
BH--WD&                                0/0 &            0/0 &      1/8    & 388/45     & 2141/128   \\
BH--NS&                                3/59 &        74/296 &    295/500  & 493/517    & 500/483    \\
BH--BH&                             10224/414&    10933/499 &  10942/494  & 10942/484  & 10939/466  \\
Total in binaries:&               28909/1326 &   28093/1809 &  23870/1947 &  18567/1840& 17747/1815 \\ 

&&&&&\\

Single: binary disruption&  16165/12439 &  27776/22161 & 34536/29391 & 35981/33260  & 36155/33913  \\
Single: binary merger&      7244/499    & 11091/1729   &18979/2802   & 24237/7519   & 24251/7575       \\  
Single progenitor&           68465/7865 & 84455/21935  &86245/32170  &86245/32170  & 86245/32170 \\  
Total single:&              91874/20803 &123322/45825  &139760/64363 &146463/72949 &146651/73658 \\ 

\enddata
\label{std03}
\tablenotetext{a}{
Same as Table~\ref{std01}.
}
\end{deluxetable}

\begin{deluxetable}{lccccc}
\tablewidth{480pt}
\tablecaption{Black Hole Populations: Retained in/Ejected from Cluster\tablenotemark{a}\ with 
              $V_{\rm esc}=300$ km s$^{-1}$ - Standard Model}
\tablehead{ Type\tablenotemark{a} & 8.7 Myr & 11.0 Myr & 15.8 Myr & 41.7 Myr & 103.8 Myr \\
 & $M_{\rm to}=25\msun$ & $M_{\rm to}=20\msun$ & $M_{\rm to}=15\msun$ & 
$M_{\rm to}=8\msun$ & $M_{\rm to}=5\msun$ }
\startdata
BH--MS&                            17948/62  & 16780/86     &  12648/88   & 6845/89    & 4192/87    \\   
BH--HG&                               14/0 &           21/0 &     19/0    & 15/0       & 14 /1     \\
BH--RG&                                0/0 &            0/0 &      0/0    &  4/0       & 9/0      \\
BH--CHeB&                           1265/0 &          1061/0&    702/0    & 286/0      & 200/0    \\ 
BH--AGB&                              25/0 &           16/0 &     20/0    & 11/0       & 14/0     \\
BH--He&                              221/0 &          136/0 &    100/0    & 288/0      & 388/0      \\
BH--WD&                                0/0 &            0/0 &      8/1    & 432/1      & 2266/3     \\
BH--NS&                               54/8  &       339/31  &    755/40   & 979/31     & 960/23     \\
BH--BH&                             10621/17&    11400/32  &  11406/30   & 11399/27   & 11382/23   \\
Total in binaries:&               30148/87   &   29753/149  &  25658/159  &  20259/148 & 19425/137  \\ 

&&&&&\\

Single: binary disruption&  25364/3240  &  43111/6826  & 53392/10535  & 56379/12882  & 56789/13279  \\
Single: binary merger&      7703/40     & 12433/387    & 21100/681    & 28682/3074   & 28739/3087       \\  
Single progenitor&           74140/2190 & 98615/7775   & 104050/14365 & 104050/14365 & 104050/14365 \\  
Total single:&             107207/5470  &154159/14988  & 178542/25581 & 189111/30321 & 189578/30731 \\ 

\enddata
\label{std04}
\tablenotetext{a}{
Same as Table~\ref{std01}.
}
\end{deluxetable}
\clearpage

\begin{deluxetable}{lccc}
\tablewidth{280pt}
\tablecaption{Retained BH Fractions for Various Models\tablenotemark{a}}  
\tablehead{   Model & $f_{\rm bin}$=0\%& $f_{\rm bin}$=50\%& $f_{\rm bin}$=100\% }
\startdata
 A:  10 $km s^{-1}$      & .53   & .40   & .27\\
 A:  50 $km s^{-1}$      & .64   & .56   & .49\\
 A: 100 $km s^{-1}$      & .73   & .69   & .64\\
 A: 300 $km s^{-1}$      & .88   & .87   & .86\\
                         &       &       &    \\
 B:  50 $km s^{-1}$      & .64   & .64   & .63\\
C1:  50 $km s^{-1}$      & .74   & .64   & .54\\
C2:  50 $km s^{-1}$      & .42   & .49   & .56\\
 D:  50 $km s^{-1}$      & .64   & .59   & .54\\
 E:  50 $km s^{-1}$      & .02   & .16   & .29\\
 F:  50 $km s^{-1}$      & .61   & .54   & .46\\
G1:  50 $km s^{-1}$      & .56   & .50   & .43\\
G2:  50 $km s^{-1}$      & .63   & .55   & .47\\
 H:  50 $km s^{-1}$      & .52   & .47   & .42\\
 I:  50 $km s^{-1}$      & .81   & .73   & .63\\
 J:  50 $km s^{-1}$      & .64   & .56   & .49\\
 
\enddata
\label{frac01}
\tablenotetext{a}{Fractions are calculated for entire BH population; both single and binary 
BHs added with the assumed initial cluster binary fraction $f_{\rm bin}$. The fractions are 
obtained for a time of 103.8  Myr after the starburst. For our (standard) Model~A we show
fractions for different escape velocities, while for all other models we assume $V_{esc} = 
50\kms$.}
\end{deluxetable}

\vspace*{10cm}
\begin{deluxetable}{lcccccc}
\tablewidth{540pt}
\tablecaption{Very Young Black Hole Populations Retained In/Ejected From Cluster With 
$V_{\rm esc} = 50$ km s$^{-1}$ for Different Models\tablenotemark{a}}
\tablehead{ Type&         A & B & C1 & C2 & D & E}
\startdata
Binaries:&&&&& & \\
BH--MS&             14055/2811  & 30786/2331  & 23471/2489 & 2702/62    & 12147/730   & 1030/2342 \\
BH--HG&                20/1     &  0/0        & 43/0       & 3/0        & 23/0       & 2/0       \\
BH--RG&                0/0      & 0/0         & 0/0        & 0/0        & 0/0       & 0/0       \\
BH--CHeB&              1057/4   &  11/0       & 1624/7     & 179/0      & 943/0       & 51/ 6     \\
BH--AGB&               16/0     & 0/0         & 19/0       &  5/0       & 14/0        & 0/0       \\
BH--He&                123/13   &  0/0        & 365/3      & 84/0       & 37/ 0       & 53/7      \\
BH--WD&                0/0      & 0/0         & 0/0        & 0/0        & 0/0         & 0/0       \\
BH--NS&               11/359    & 0/3         & 41/630     & 67/30      & 11/127      & 0/192     \\
BH--BH&              10410/1022 &   24/3      & 22120/1121 & 4098/21    & 10149/1401  & 43/272    \\
Total:&              25692/4210 & 30821/2337  & 47683/4250 & 7138/113   & 23324/2258  & 1179/2819 \\
&&&&& & \\

Single:&&&&& & \\
binary disruption&  18136/31801& 5067/14023 & 7317/16957   &27101/30609 & 17544/23045 &15978/75191 \\
binary merger&    9906  /2914  &5270 /1881  &  8246/2243   & 9423 /1654 & 10232/2123 & 8412 /2732  \\
Single progenitor& 75465/30925 & 75465/30925& 97745/17350  &42580/59425 &75465/30925& 1820/105340\\
Total:&          103507 /65640 & 85802/46829& 113308/36550 &79104/91688 &103241/56093 &26210/183263 \\
&&&&&& \\
\hline
\hline
&&&&& & \\
Type&         F & G1 & G2 & H & I & J \\

\hline
&&&&& & \\
Binaries:&&&&& & \\
BH--MS&        8046/1682    & 10925/2721 & 14692/2921  & 14203/4110 & 11346/9 & 13960/983       \\      
BH--HG&        10/0         & 24/0       & 32/1        & 20/1       & 17/0    & 11/0      \\
BH--RG&        0/0          & 0/0        & 0/0         & 0/0        & 0/0     & 0/0       \\
BH--CHeB&      594/6        & 747/6      &1056/6       & 1089/4     & 919/0   & 1061/0       \\
BH--AGB&       7/0          &14/0        & 12/0        & 16/0       & 16/0    & 23/0       \\
BH--He&        87/5         & 105/8      & 128/13      & 126/19     & 19/0    & 86/3       \\
BH--WD&        0/0          & 0/0        & 0/0         & 0/0        & 0/0     & 0/0       \\
BH--NS&        0/207        & 4/324      &  7/404      & 1/75       & 0/ 0    & 9/265       \\
BH--BH&        4858/552     & 2885/634   & 8719/1024   & 10420/1353 &10085/477& 10568/927        \\
Total:&        13602/2452   & 14704/3693 &24646 /4369  & 25875/5562 &22402/486  &25718  /2178      \\
&&&&& & \\

Single:&&&&& & \\
binary disruption&10272/18773&14656/27049& 19631/34003 & 18582/36967&10343/11915 & 20599/36724       \\
binary merger&    4325/1233 & 3132/1728  &7816  /2625  & 10413/3329 &  8720/1151 &  8823 /1592      \\ 
Single progenitor&43870/19615&54695/30545& 71515/31090 & 75465/30925  & 68845/16640& 75465/30925                 \\ 
Total:&        58467/39621  &72483/59322 &98962 /67718 & 104460/71221 & 87908/29706&104887/69241     \\
 
\enddata
\label{parm01}
\tablenotetext{a}{All numbers correspond to an age of 11 Myrs ($M_{\rm to}=20 M_\odot$)}
\end{deluxetable}

\begin{deluxetable}{lcccccccccccc}
\tablewidth{540pt}
\tablecaption{Young Black Hole Populations Retained In/Ejected From Cluster 
With $V_{\rm esc} = 50$ km s$^{-1}$ for Different Models\tablenotemark{a}}
\tablehead{ Type&         A & B & C1 & C2 & D & E}
\startdata
Binaries:&&&&& & \\
BH--MS&             3317/962    &30116/2261  & 5448/774    & 816/45     & 3039/98   & 72/789    \\
BH--HG&               11/4      &   6/1      & 26/4        & 1/1        &  4/1      & 0/9       \\
BH--RG&                4/5      &  1/0       & 0/2         & 0/0        & 0/0       & 0/2       \\
BH--CHeB&              187/13   &   43/1     & 285/4       & 59/0       & 156/5     & 2/13      \\
BH--AGB&               14/0     &   4/0      & 24/0        & 0/0        & 11/0      & 0/0    \\
BH--He&                 45/343  &  11/29     & 203/292     & 3/3        & 59/226    & 17/259    \\
BH--WD&             1682/587    &  235/36    & 3119/623    & 491/7      & 1431/139  & 201/506   \\
BH--NS&              181/802    &   5/7      & 651/1498    &113 /48     & 95/236    & 45/415    \\
BH--BH&              10415/990  &   24/3     & 22068/974   & 4097/20    &10146/970  & 44/266    \\
Total:&              15856/3706 & 30445/2338 & 31824/4171  & 5580/124   &14941/1675 & 381/2259  \\
&&&&& & \\

Single:&&&&& & \\
binary disruption&  22900/47168&  5509/18263 & 12414/39071 &29358/33357 &22179/37215&19871/83178 \\
binary merger&      20530/11096& 11740/6853  & 18390/10512 &20127/10275 &27122/15666& 19417/10535 \\
Single progenitor&  75830/42585& 75830/42585 & 92325/32465 &42580/59425 &75830/42585& 2010/117555           \\
Total:&            119360/100949& 93079/67701 &123129/82048 & 92065/103057&125131/95466 &41298/221268 \\

&&&&&& \\
\hline
\hline
&&&&& & \\
Type&         F & G1 & G2 & H & I & J \\

\hline
&&&&& & \\
Binaries:&&&&& & \\
BH--MS&      1890/578        & 2522/938    & 3483/979    & 3319/1482  & 2952/8    & 3493/622       \\
BH--HG&      1/6             &2 / 9        & 3/11        &11/5        &10/0       & 7/2      \\
BH--RG&      0/4             & 1/4         & 0/5         & 4/6        & 3/0       & 2/0       \\
BH--CHeB&   103/8            &127/5        & 173/14      & 187/13     & 181/2     & 188/8       \\
BH--AGB&     8/0             &10/0         & 12/0        & 14/0       & 14/0      & 11/0       \\
BH--He&      32/215          & 38/325      & 63/362      & 45/378     & 1/27      & 18/50      \\
BH--WD&      961/308         & 1401/560    & 1879/577    & 1743/629   & 1094/9    & 1609/199  \\
BH--NS&      87 /502         & 132/738     & 185/929     & 165/755    & 40/21     & 129/718     \\
BH--BH&      4866/537        & 2898/613    & 8732/1000   & 10441/1434 & 10103/502 & 10578/896    \\
Total:&      7948/2158       &7131 /3192   & 14530/3877  & 15929/4702 & 14398/569 & 16035/2495       \\
&&&&& & \\

Single:&&&&& & \\
binary disruption&13547/28403& 19435/40297 & 24946/50147 & 28917/72295&12821/20559 & 25774/52599       \\
binary merger& 11225/7522    & 13100/9913  & 18325/11120 & 21892/13710 & 16923/4345 & 16475/6738      \\ 
Single progenitor&44115/28245& 55075/42620 & 71870/42440 & 76280/70605 & 68845/16640&  75830/42585                  \\ 
Total:&       68887/64170  &   87610/92830 &115141/103707&127089/156610& 98589/41544&118079/101922       \\

\enddata
\label{parm02}
\tablenotetext{a}{All numbers correspond to an age of 103.8 Myrs ($M_{\rm to}=5 M_\odot$)}
\end{deluxetable}
\clearpage

\begin{figure}
\includegraphics[width=0.9\columnwidth]{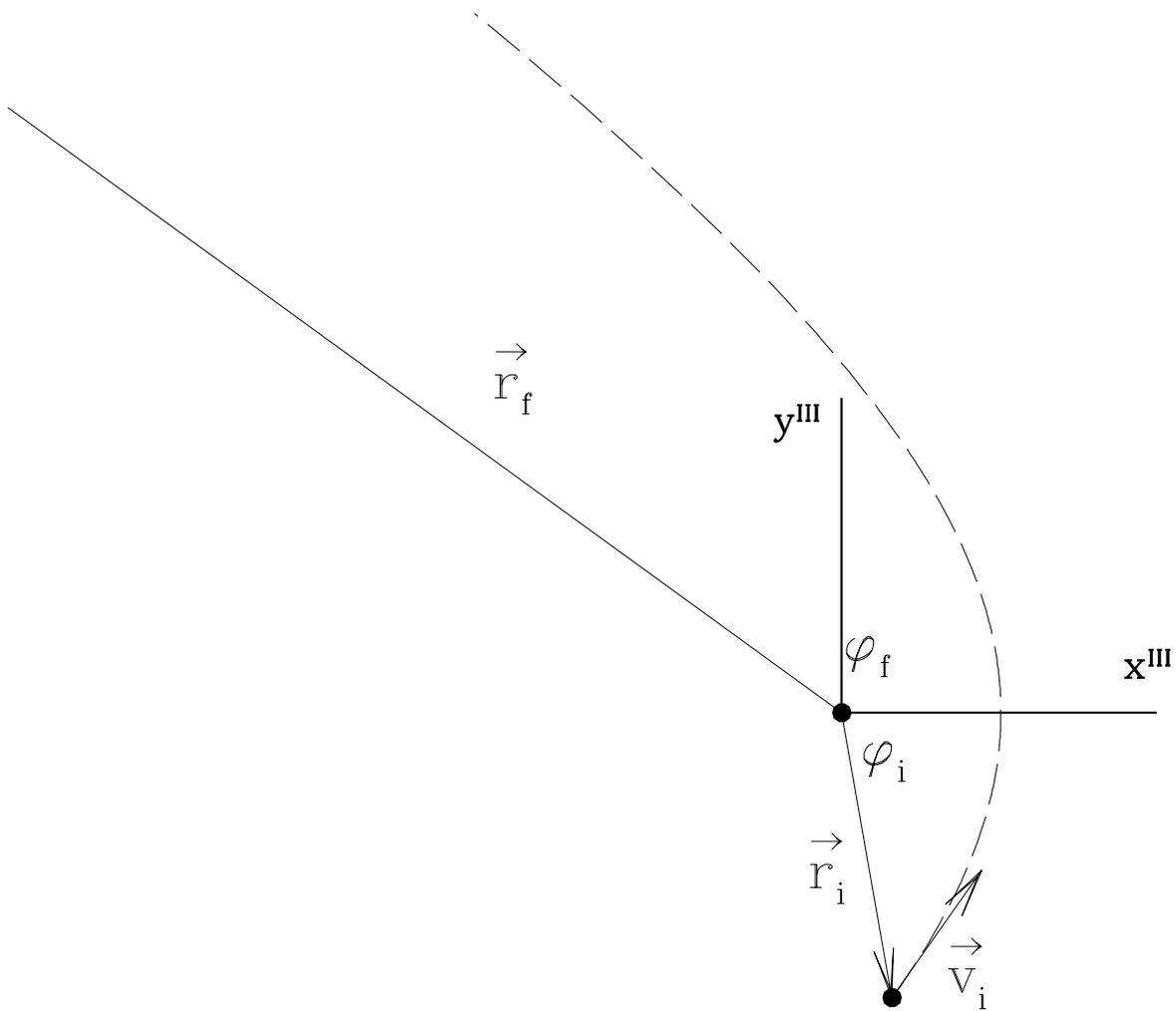}
\caption{
The trajectory of the newly born compact object in the system
connected with the companion. For details see \S\,2.3.
}
\label{fig02}
\end{figure}
\clearpage

\begin{figure}
\includegraphics[width=0.9\columnwidth]{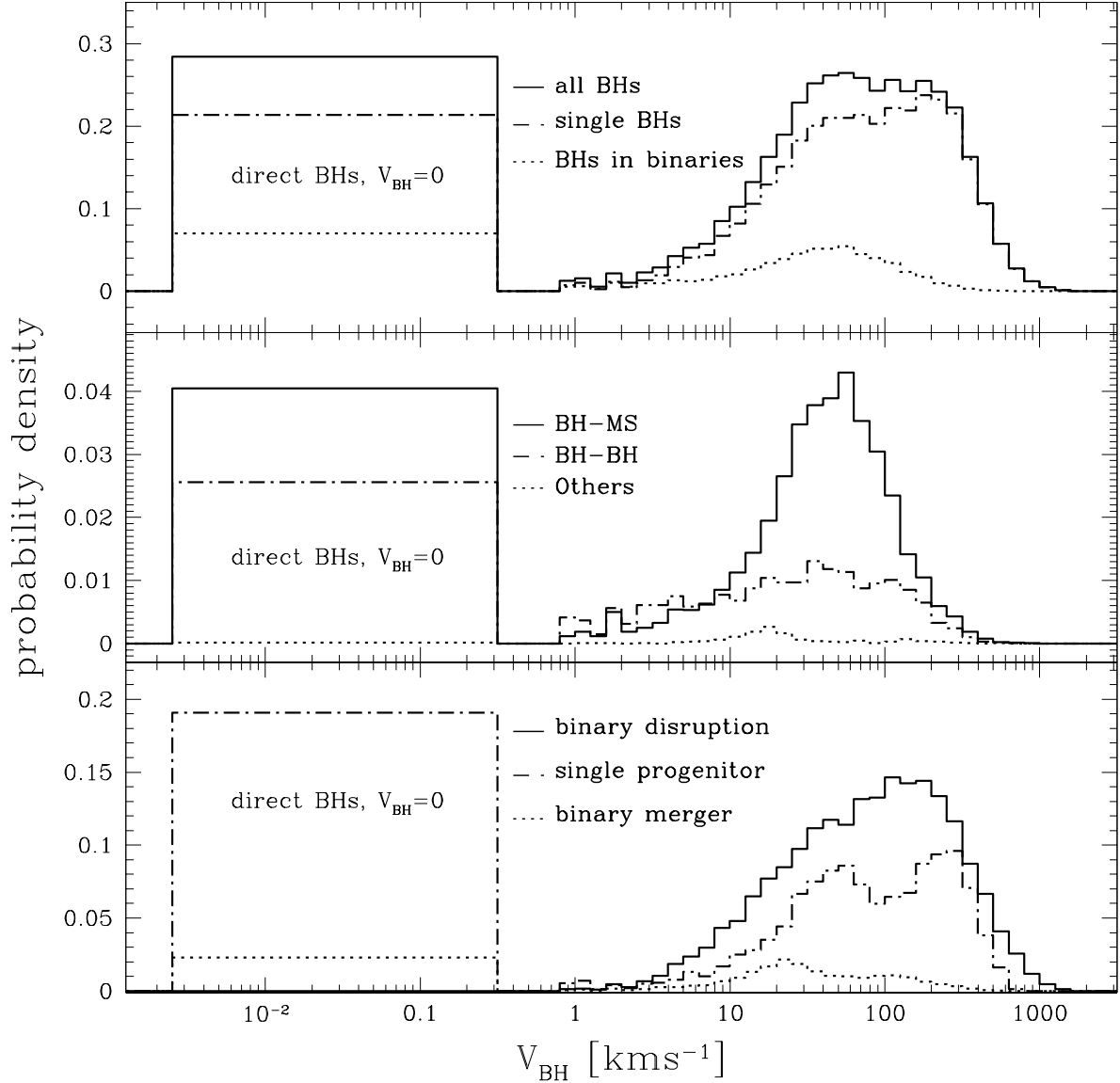}
\caption{
Spatial velocities of black holes at $t=8.7\,$Myr after the starburst.
Overall distributions (all, single, binary BHs) are shown in the top panel. 
Middle panel shows various binary BH systems, while bottom panel shows the 
single BH populations. Note generally higher systemic speeds for single BHs. 
All curves are normalized to the total number of BHs at a given time (the 
distributions show $\,dN/\,d\log V_{\rm BH}$). The no-kick  
BHs (direct formation) are contained in the left rectangular area. The area of 
the rectangle illustrates  the relative numbers of no-kick and high-velocity systems.}
\label{t1vel}
\end{figure}
\clearpage

\begin{figure}
\includegraphics[width=0.9\columnwidth]{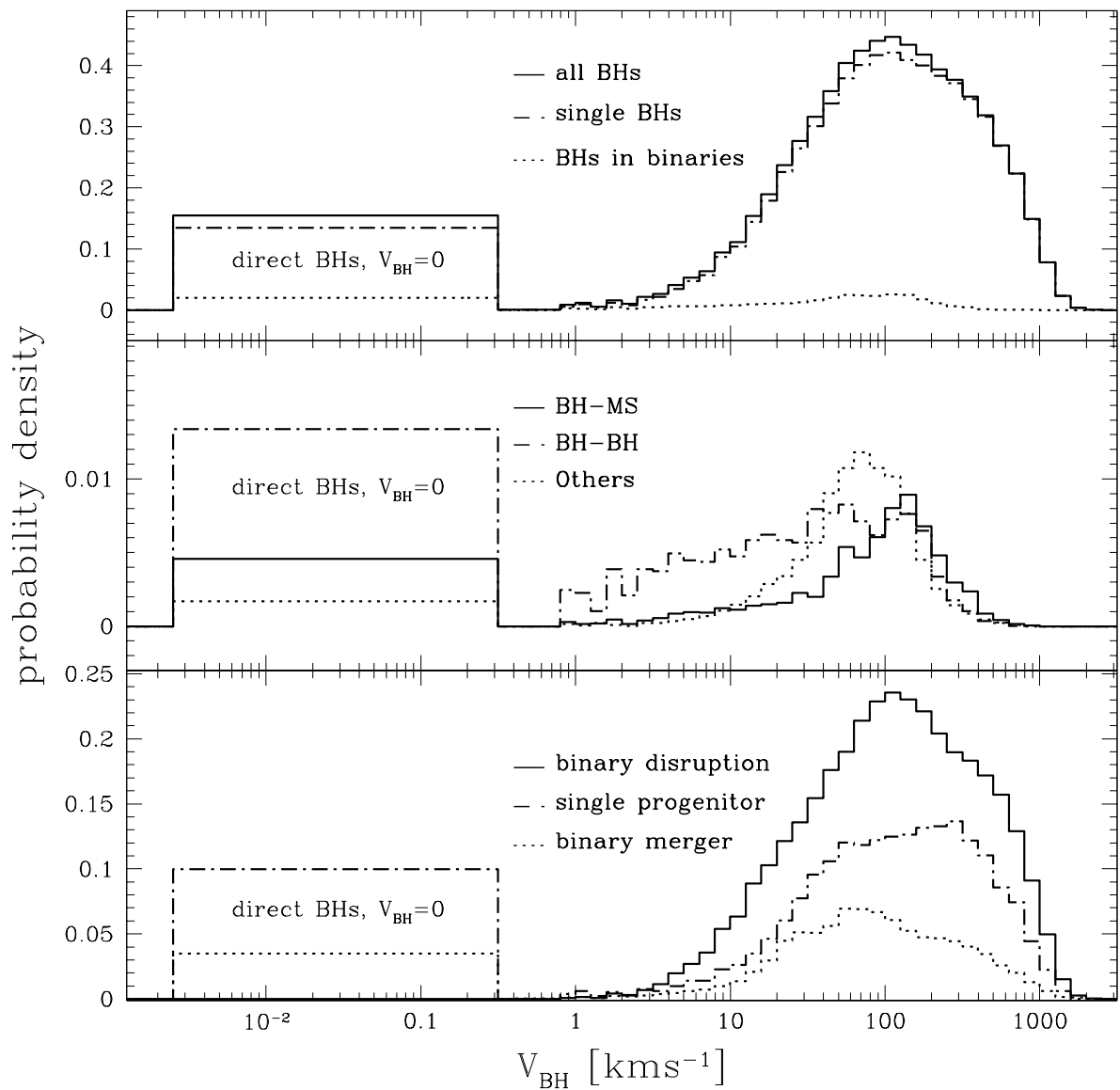}
\caption{
Spatial velocities of black holes at $t=103.8\,$Myr after the starburst. Lines 
same as for Fig.~\ref{t1vel}.
}
\label{t5vel}
\end{figure}
\clearpage

\begin{figure}
\includegraphics[width=0.9\columnwidth]{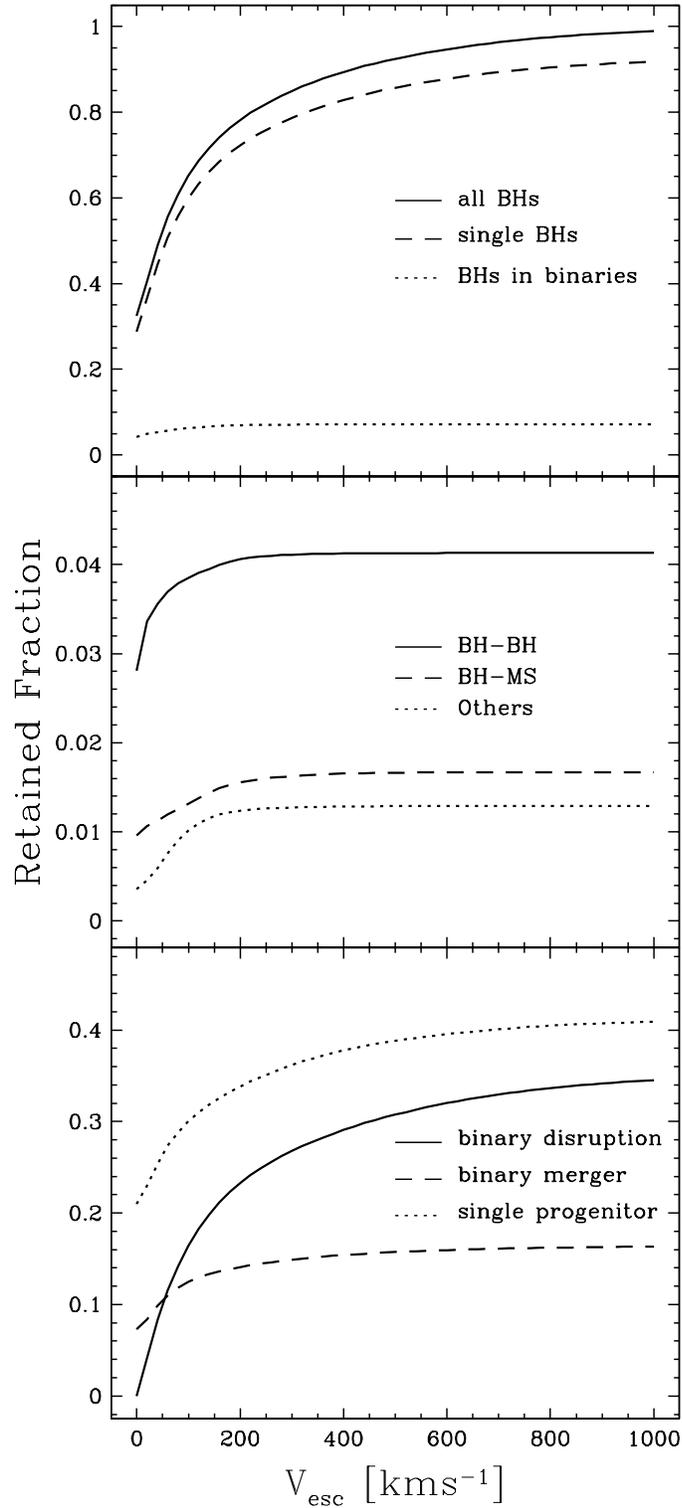}
\caption{Retained fraction (cluster population) of BHs as a function of 
$V_{esc}$ for our standard model at $t=103.8\,$Myr. Top panel shows overall 
population with contributions of single and binary BHs. Middle and bottom 
panels show various subpopulations of binary and single BHs, respectively.
All curves are normalized to total number of BHs (single and binaries) formed 
in the standard model simulation. 
}
\label{retained.all}
\end{figure}
\clearpage

\begin{figure}
\includegraphics[width=0.9\columnwidth]{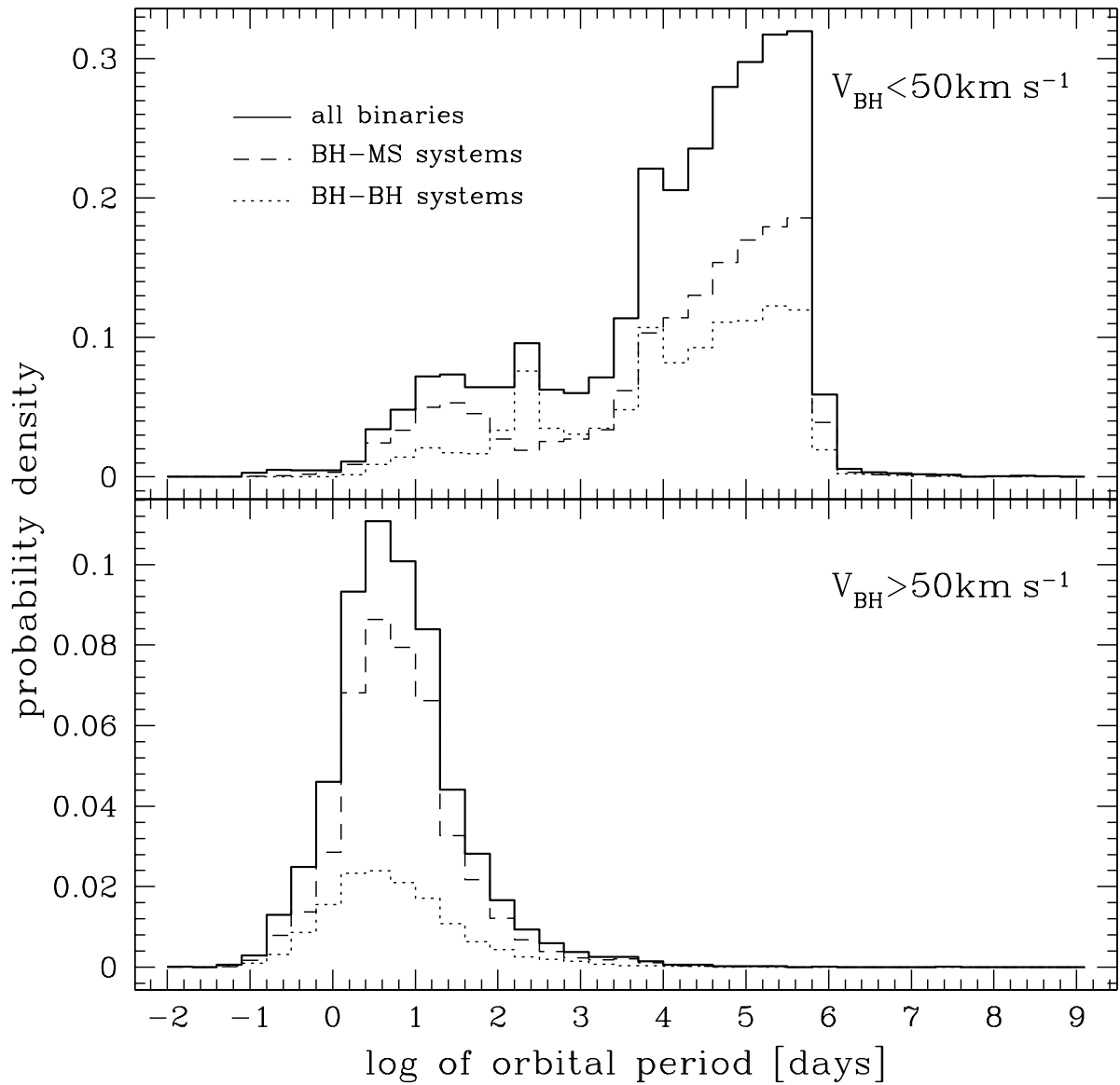}
\caption{
Period distribution of BH binaries retained in/ejected from cluster with 
$V_{\rm esc}=50$ km s$^{-1}$ at $11\,$Myr for standard model.
Two major contributing system types are shown separately: BH--MS binaries
(dashed line) and  BH--BH binaries (dotted line). All curves are 
normalized to total number of BHs (single and binaries). The distributions
show $\,dN/\,d\log P_{\rm orb}$.  
Note the different vertical scale on the panels.
}
\label{A.50.per}
\end{figure}
\clearpage

\begin{figure}
\includegraphics[width=0.9\columnwidth]{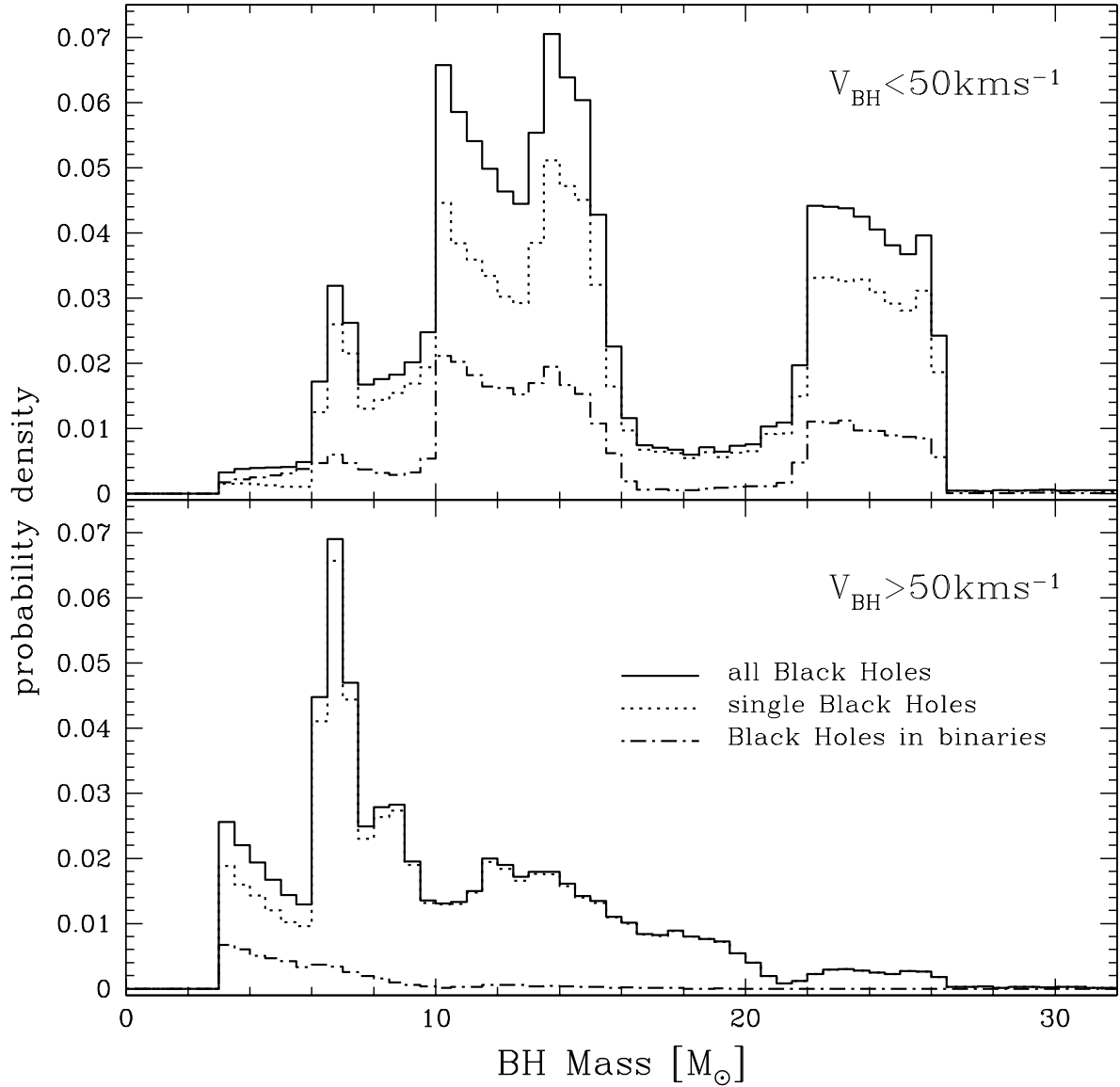}
\caption{
Mass distribution of BHs retained in/ejected from cluster with
$V_{\rm esc}=50$ km s$^{-1}$ at $11\,$Myr for standard model.
Overall distribution is shown with a solid line, while single BHs are 
shown with a dotted line and BHs in binaries with a dashed line.
All curves are normalized to total number of BHs (single and binaries).
}
\label{A.50.mass.all}
\end{figure}
\clearpage

\begin{figure}
\includegraphics[width=0.9\columnwidth]{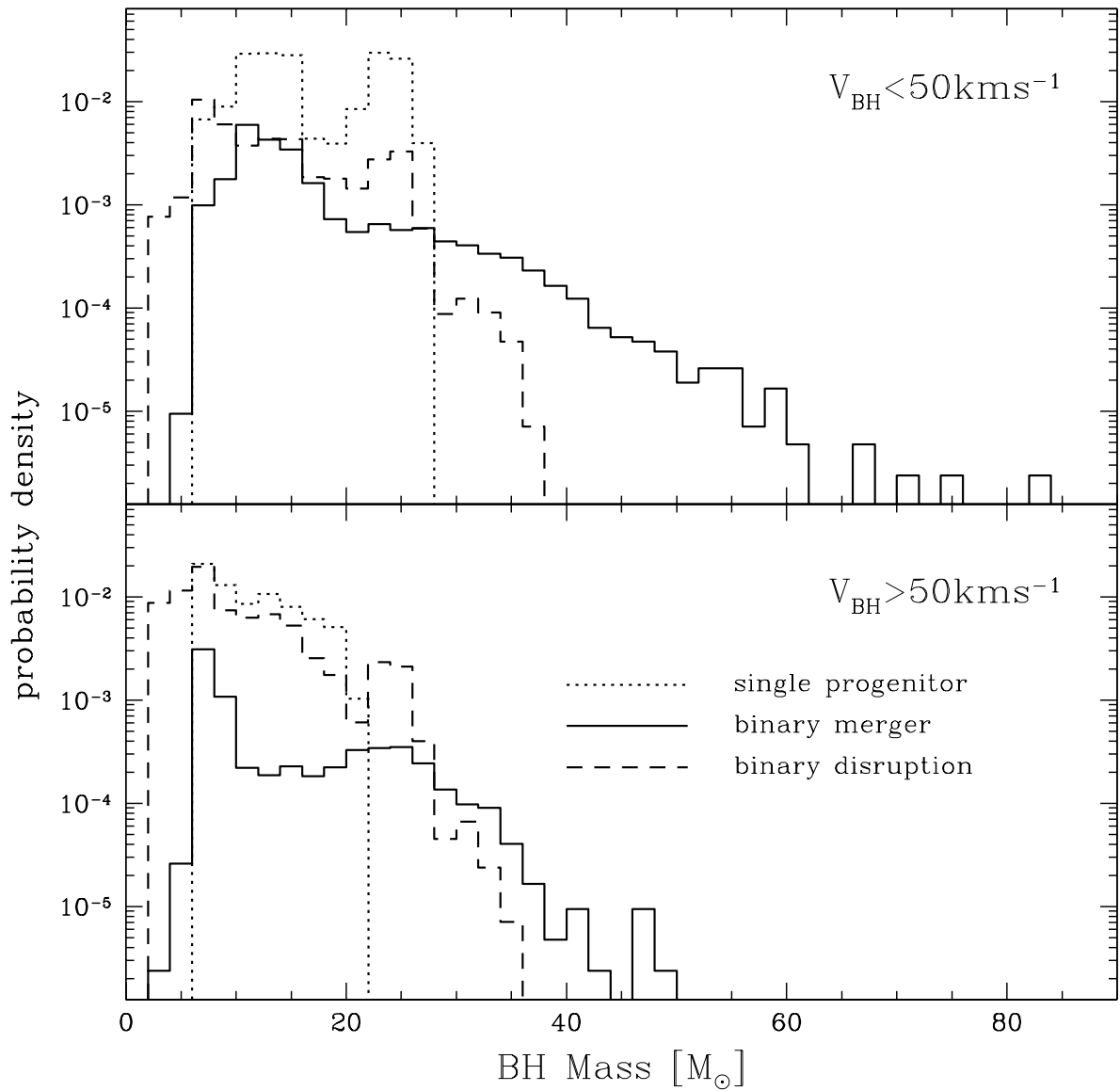}
\caption{
Mass distributions of various kinds of single BHs retained in/ejected 
from cluster with $V_{\rm esc}=50$ km s$^{-1}$ at $11\,$Myr for standard 
model.
The dotted line shows BHs originating from primordial single stars; the
dashed line represents single BHs from disrupted binaries; the
solid line is for single BHs that are remnants of merged binaries.
All curves are normalized to total number of BHs (single and binaries).
Note that, in contrast to Fig.~\ref{A.50.mass.all}, a logarithmic scale 
is used and the entire range of BH masses is shown.
}
\label{A.50.mass.sin}
\end{figure}
\clearpage

\begin{figure}
\includegraphics[width=0.9\columnwidth]{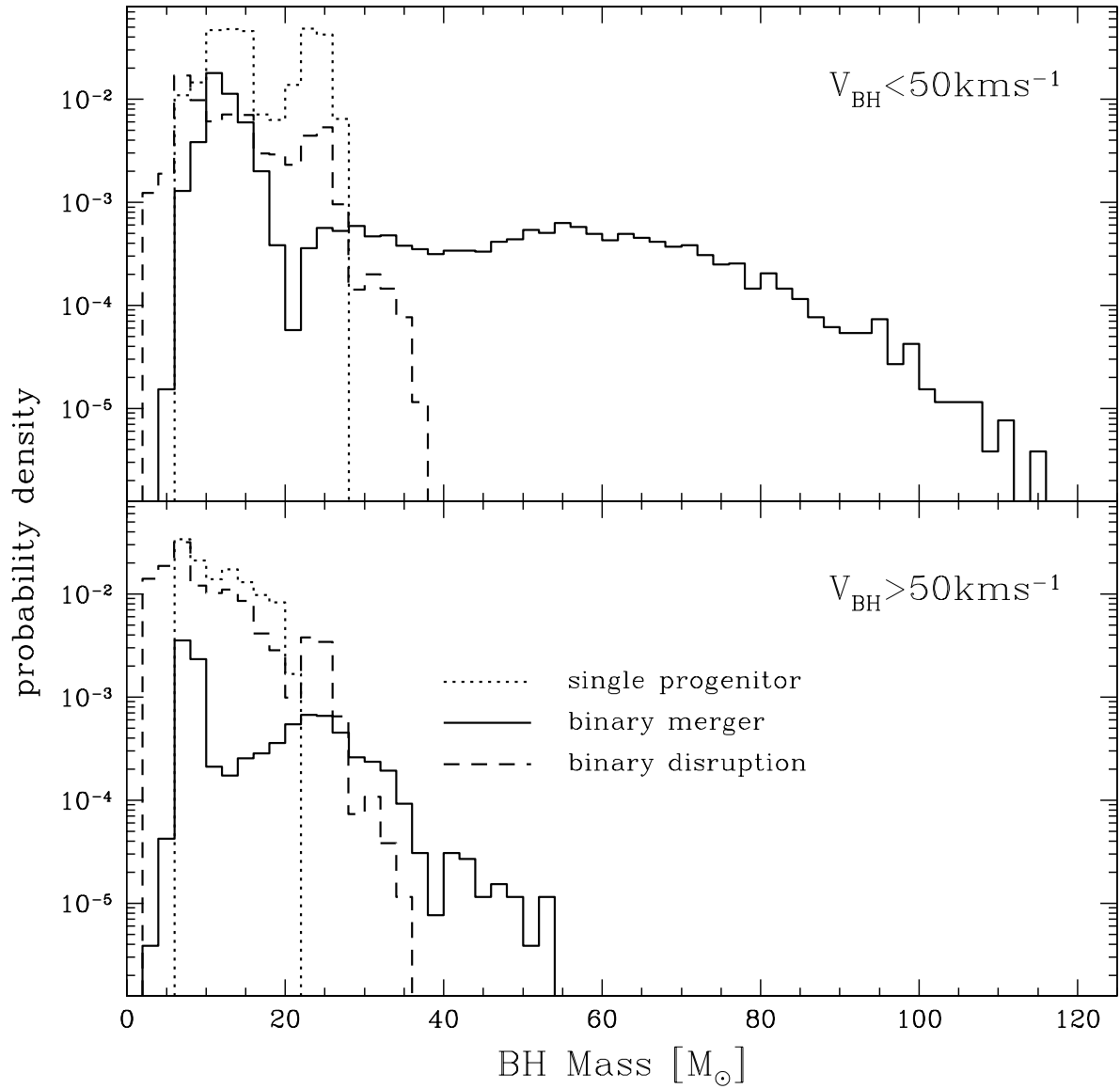}
\caption{
Same as Fig.~\ref{A.50.mass.sin} but for model in which the merger mass is calculated 
from the total mass of two merging binary components. See \S\,3.1.3  for details.
}
\label{A.50.fullmass.sin}
\end{figure}
\clearpage

\begin{figure}
\includegraphics[width=0.9\columnwidth]{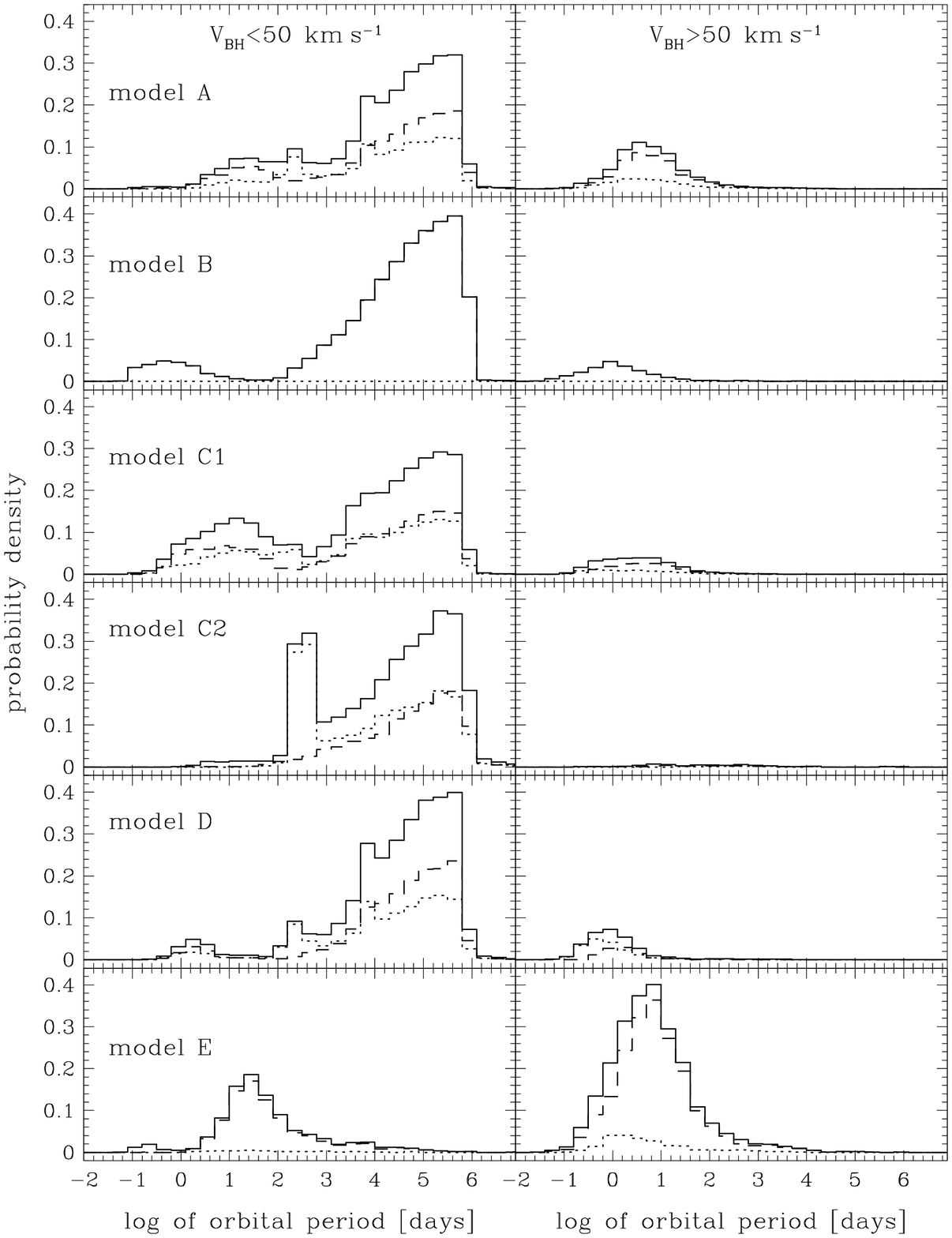}
\caption{
Period distribution of BH binaries retained in/ejected from cluster with
$V_{\rm esc}=50$ km s$^{-1}$ at $11\,$Myr for models A -- E.
Notations are the same as in Fig.~\ref{A.50.per}.
}
\label{mod.per1}
\end{figure}
\clearpage

\begin{figure}
\includegraphics[width=0.9\columnwidth]{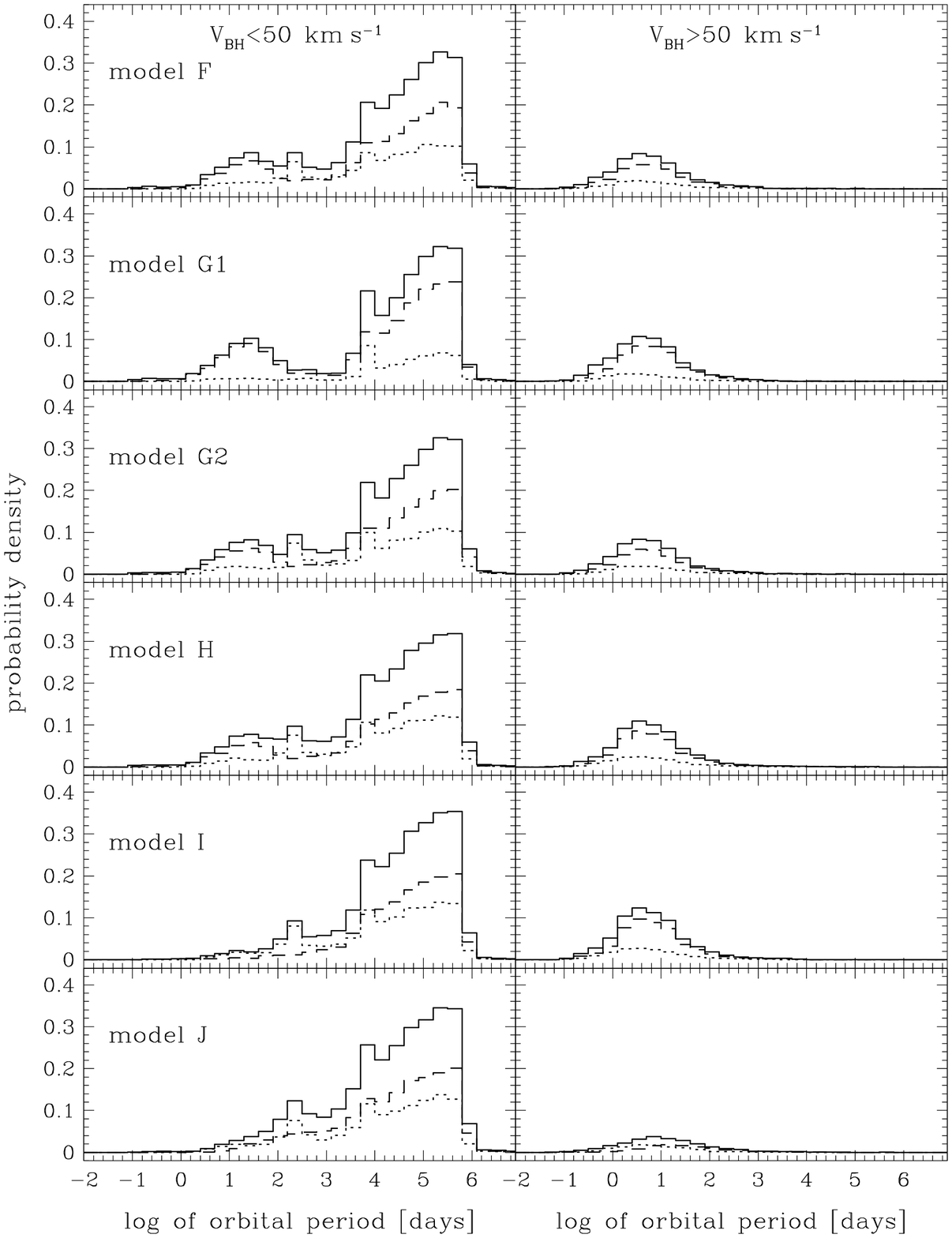}
\caption{
Period distribution of BH binaries retained in/ejected from cluster with
$V_{\rm esc}=50$ km s$^{-1}$ at $11\,$Myr for models F -- J.
Notations are the same as in Fig.~\ref{A.50.per}.
}
\label{mod.per2}
\end{figure}
\clearpage

\begin{figure}
\includegraphics[width=0.9\columnwidth]{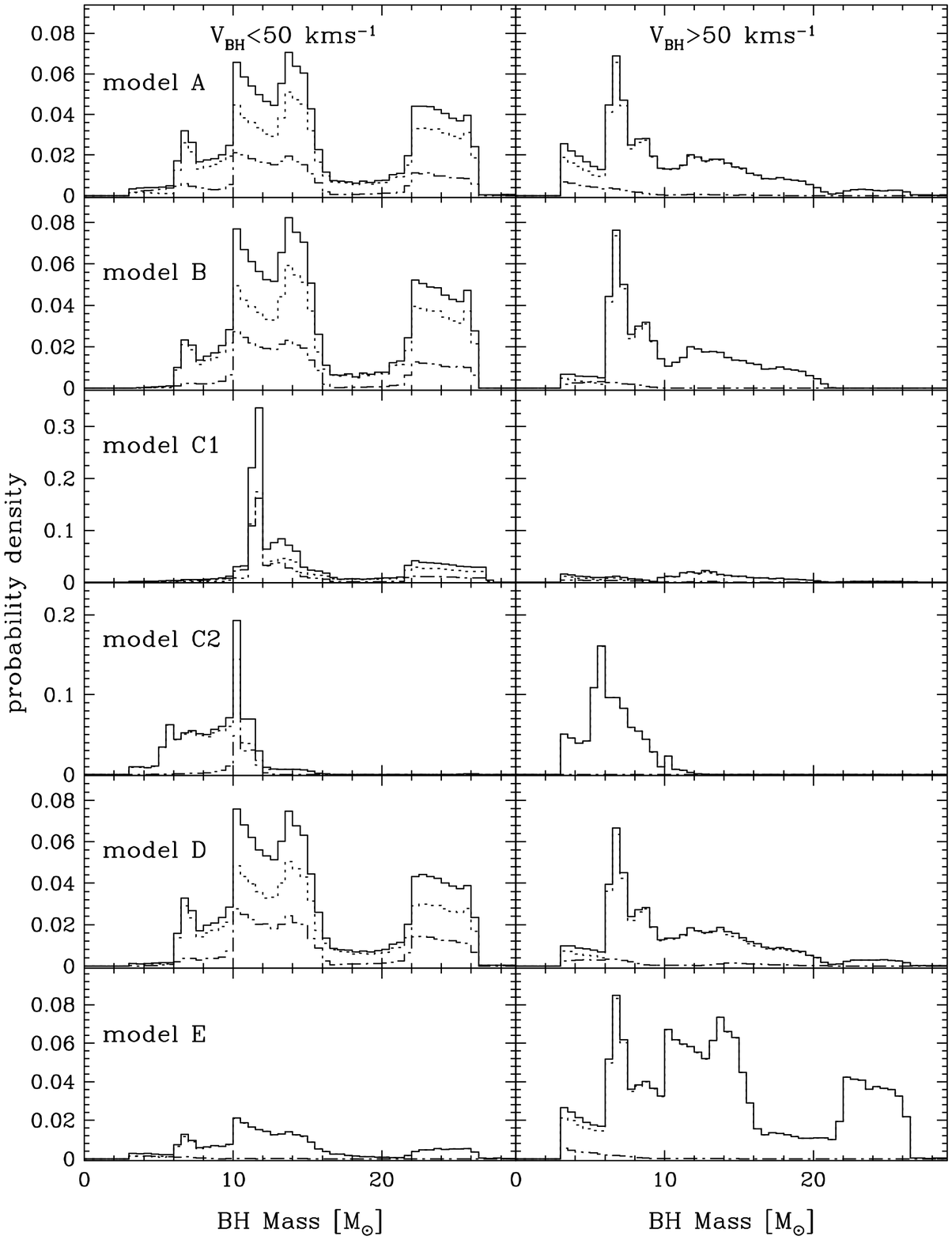}
\caption{
Mass distribution of BHs retained in/ejected from cluster with
$V_{\rm esc}=50$ km s$^{-1}$ at $11\,$Myr for models A -- E.
Notations are the same as in Fig.~\ref{A.50.mass.all}.
}
\label{mod.mas.all1}
\end{figure}
\clearpage

\begin{figure}
\includegraphics[width=0.9\columnwidth]{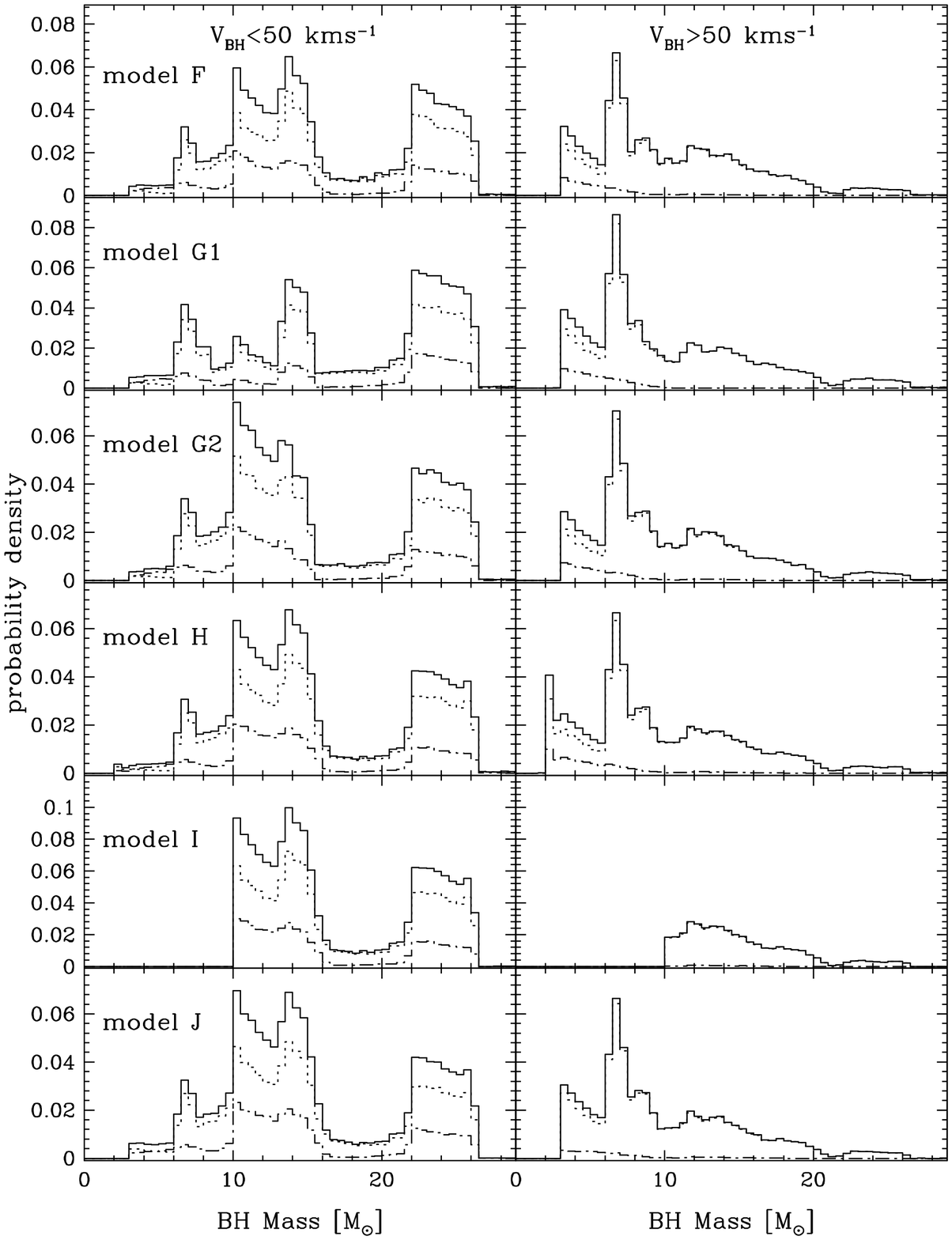}
\caption{
Mass distribution of BHs retained in/ejected from cluster with
$V_{\rm esc}=50$ km s$^{-1}$ at $11\,$Myr for models F -- J.
Notations are the same as in Fig.~\ref{A.50.mass.all}.
}
\label{mod.mas.all2}
\end{figure}
\clearpage

\begin{figure}
\includegraphics[width=0.9\columnwidth]{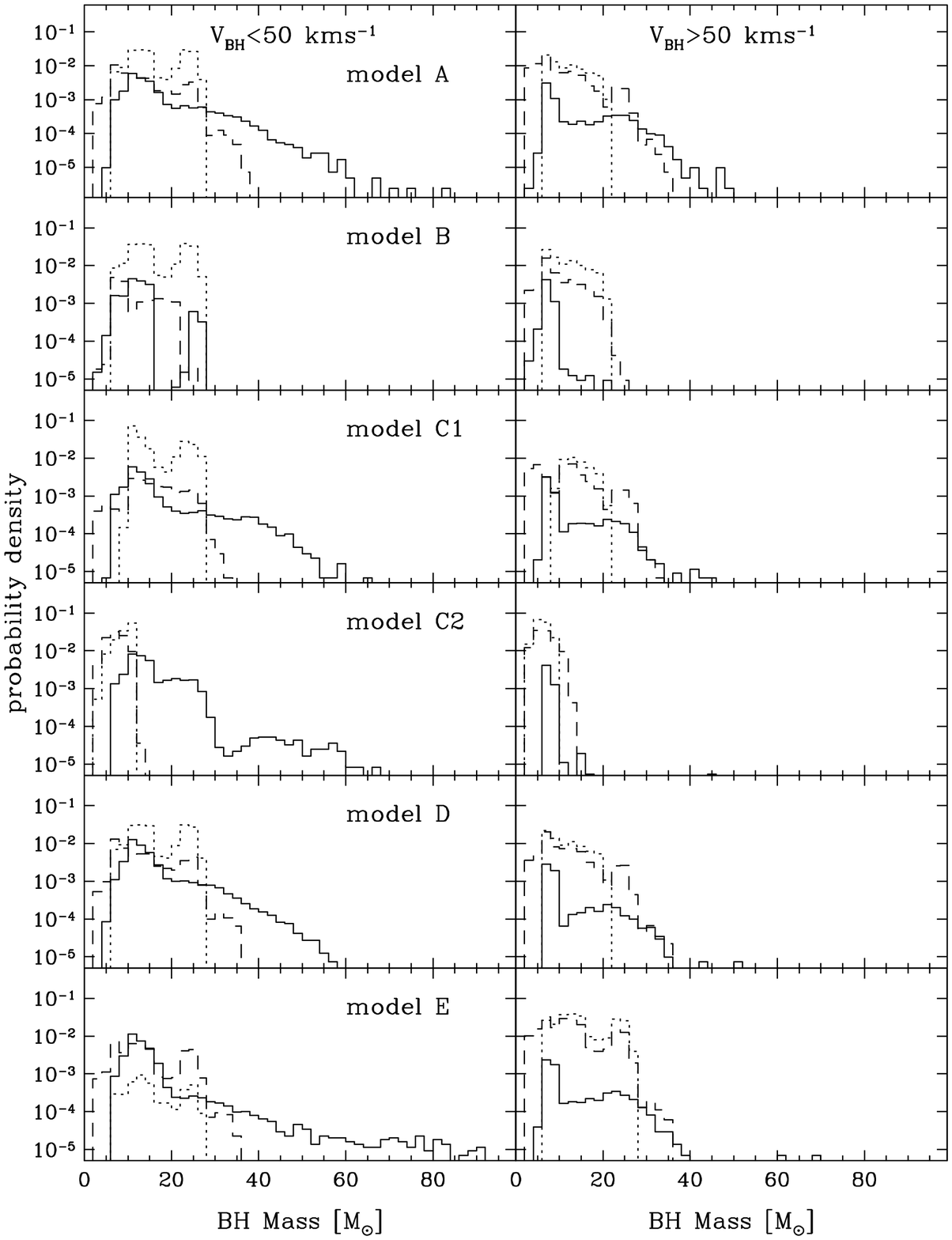}
\caption{
Mass distributions of various kinds of single BHs retained in/ejected   
from cluster with $V_{\rm esc}=50$ km s$^{-1}$ at $11\,$Myr for models A -- E. 
Notations are the same as in Fig.~\ref{A.50.mass.sin}.
}
\label{mod.mas.sin1}
\end{figure}
\clearpage

\begin{figure}
\includegraphics[width=0.9\columnwidth]{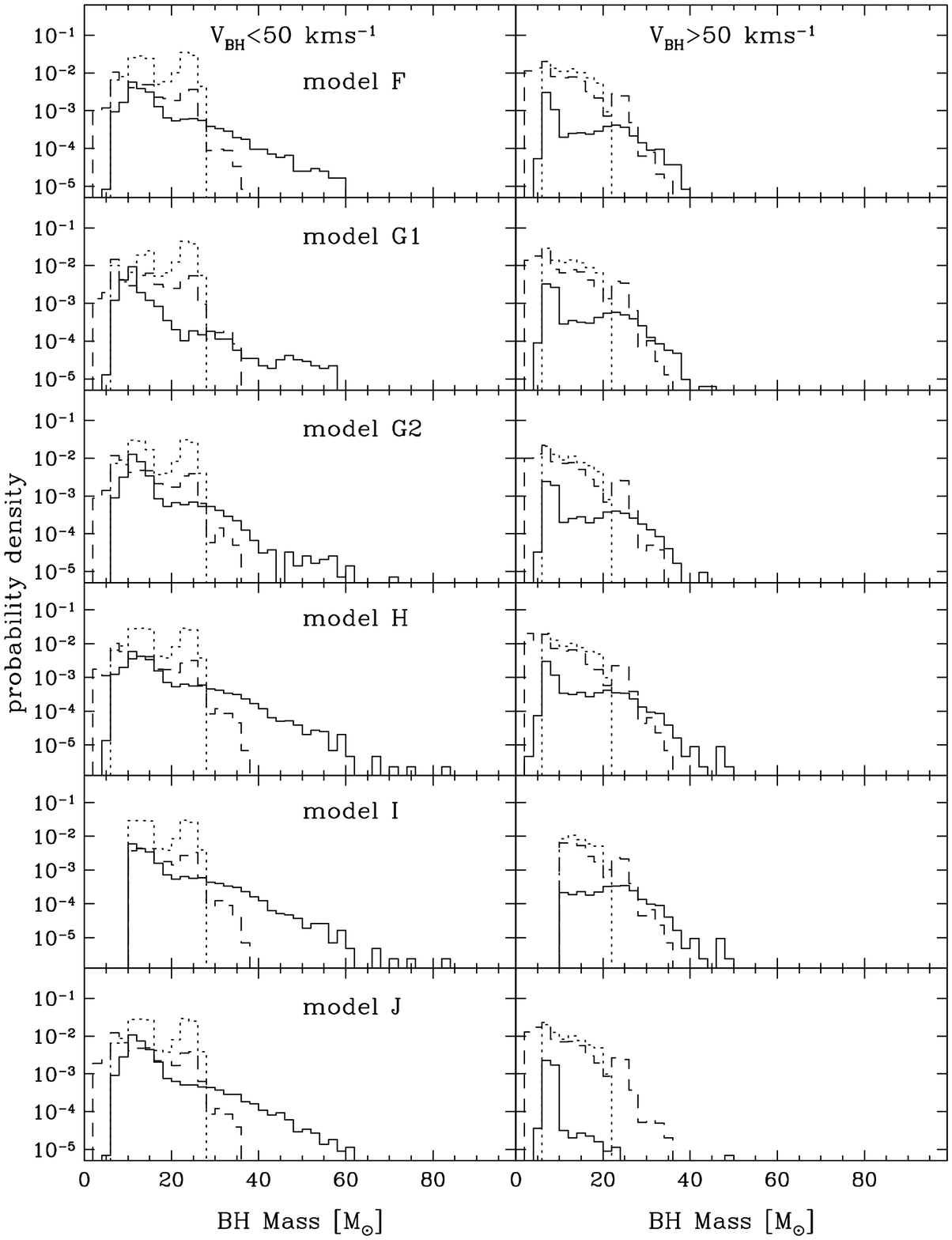}
\caption{
Mass distributions of various kinds of single BHs retained in/ejected   
from cluster with $V_{\rm esc}=50$ km s$^{-1}$ at $11\,$Myr for models F -- J.  
Notations are the same as in Fig.~\ref{A.50.mass.sin}.
}
\label{mod.mas.sin2}
\end{figure}
\clearpage

\begin{figure}
\includegraphics[width=0.9\columnwidth]{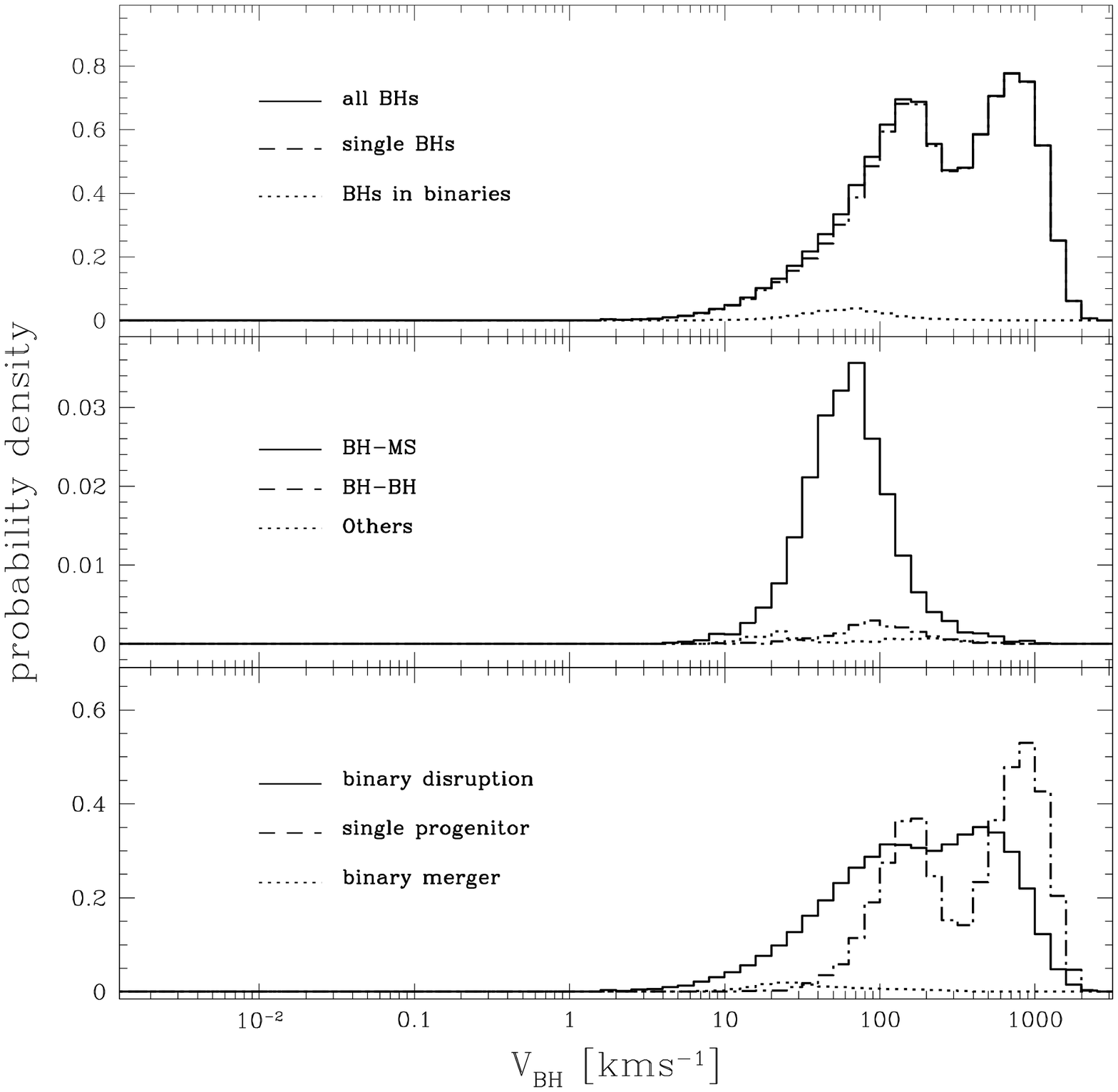}
\caption{
Spatial velocities of black holes at $t=8.7\,$Myr after the starburst 
for Model E. Lines same as for Fig.~\ref{t1vel}.
Note absence of the no-kick BHs (cf.\ Fig.~\ref{t1vel}). 
}
\label{t1velE}
\end{figure}
\clearpage

\begin{figure}
\includegraphics[width=0.9\columnwidth]{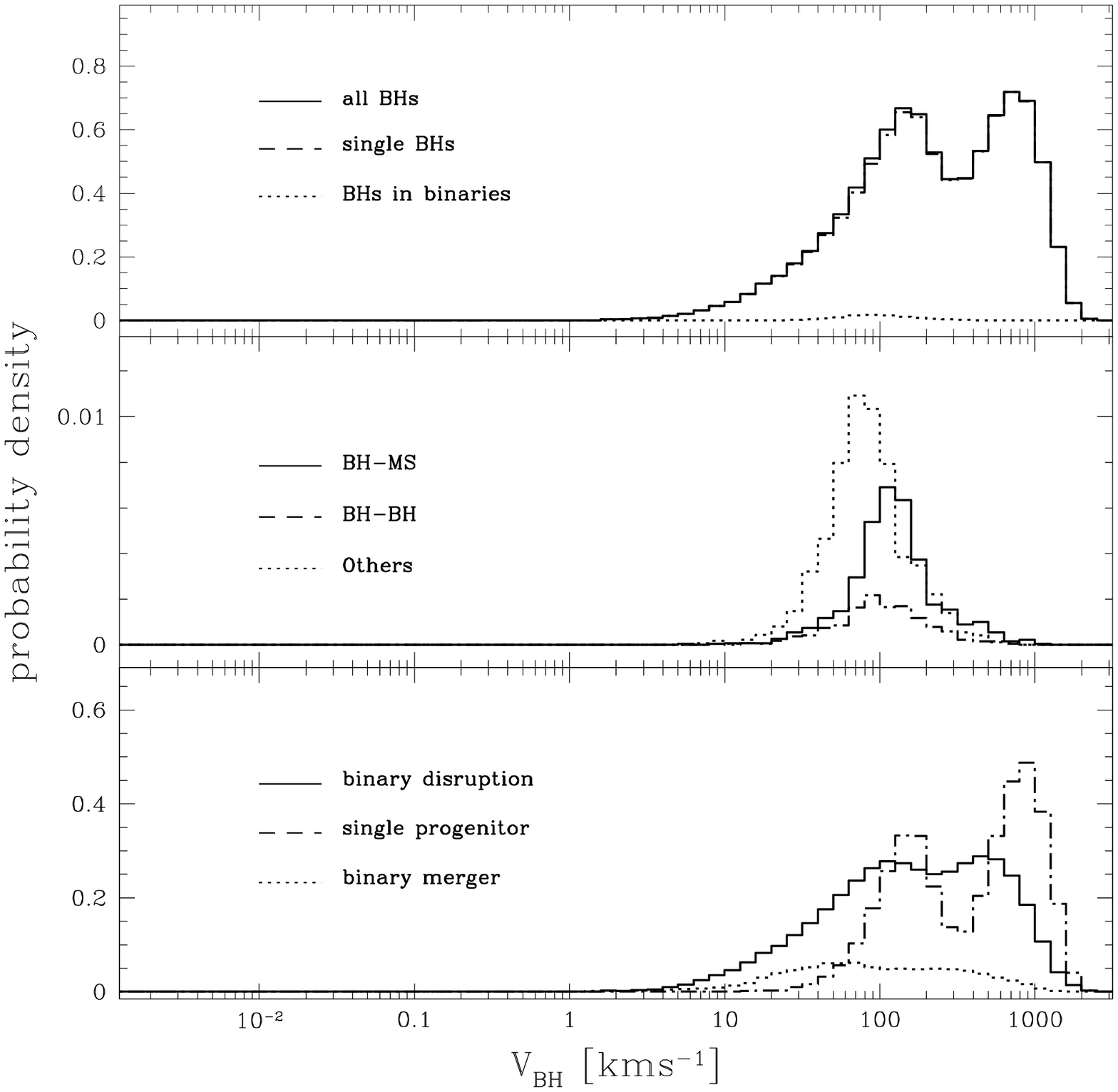}
\caption{
Spatial velocities of black holes at $t=103.8\,$Myr after the starburst 
for Model E. Lines same as for Fig.~\ref{t1vel}.
Note absence of the no-kick BHs (cf.\ Fig.~\ref{t1vel}).
}
\label{t5velE}
\end{figure}
\clearpage

\begin{figure}
\includegraphics[width=0.9\columnwidth]{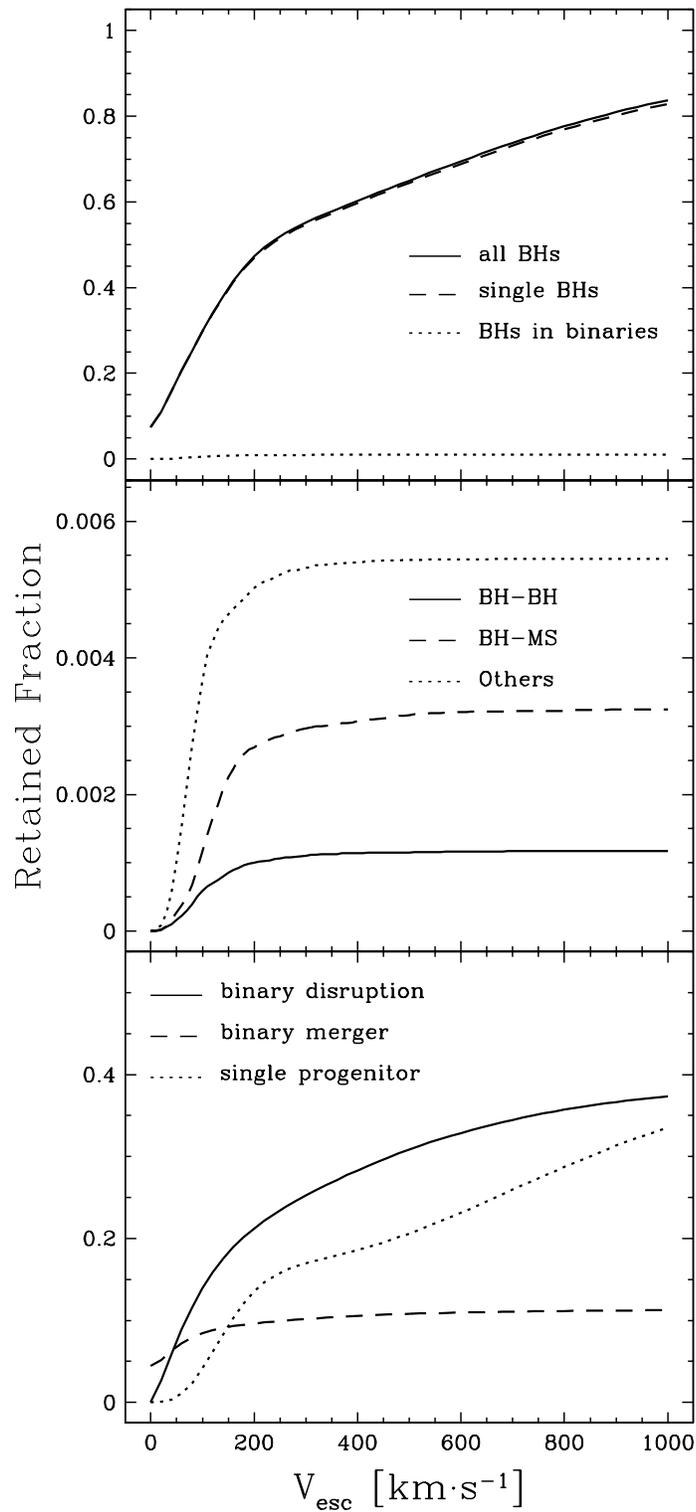}
\caption{Retained fraction (cluster population) of BHs as a function of 
$V_{esc}$ for model E and $t=103.8\,$Myr. Line styles are the same as for
Fig.~\ref{retained.all}. All curves are normalized to the total number of BHs
(single and binaries) formed in the standard model simulation. Note that the
fraction showing all BHs does not reach unity, since there is still a 
small number of BHs with velocities over 1000 km s$^{-1}$.
}
\label{retained.allE}
\end{figure}
\clearpage

\end{document}